\documentclass[twocolumn]{aastex63}

\usepackage{multirow}
\usepackage{amsmath}
\usepackage{longtable}
\usepackage{color}
\usepackage{lineno}

\newcommand{\vlsr}     {V_\mathrm{lsr}}
\newcommand{\vsys}     {V_\mathrm{sys}}

\newcommand{\pa} {\mathrm{P.A.}}
\newcommand{\vcb}     {v_\mathrm{CB}}
\newcommand{\rcb}     {r_\mathrm{CB}}

\newcommand{\Hii}{H{\sc ii}}

\newcommand{\h} {^{\mathrm{h}}}
\newcommand{\m} {^{\mathrm{m}}}
\newcommand{\s} {^{\mathrm{s}}}
\newcommand{\Jybeam}  {\mbox{Jy}~\mbox{beam}^{-1}}
\newcommand{\mJybeam}  {\mbox{mJy}~\mbox{beam}^{-1}}

\newcommand{\cm}	{\mbox{cm}}

\newcommand{\kms}	{\mbox{km s}^{-1}}

\newcommand{\K}	{{\rm K}}
\newcommand{\gcm}	{{\rm g}~{\rm cm}^{-2}}
\newcommand{\au} {\mbox{au}}
\newcommand{\dv}       {\Delta v}
\newcommand{\dnu}     {\Delta \nu}
\newcommand{\tauff}     {\tau_\mathrm{ff}}
\newcommand{\tauhrl}     {\tau_\mathrm{HRL}}
\newcommand{\Te}         {T_e}
\newcommand{\Tff}         {T_\mathrm{ff}}
\newcommand{\Thrl}         {T_\mathrm{HRL}}
\newcommand{\elecm}     {\mathrm{EM}}
\newcommand{\pccm}  {\mbox{pc cm}^{-6}}
\newcommand{\GHz} {\mbox{GHz}}

\newcommand{\kHz} {\mbox{kHz}}
\newcommand{\Jy} {\mbox{Jy}}
\newcommand{\mJy} {\mbox{mJy}}
\newcommand{\kpc} {\mbox{kpc}}
\newcommand{\chisq}{\chi^2}

\shorttitle{A Multi-Scale ALMA View of G35.20-0.74N}
\shortauthors{Zhang et al.}

\begin{document}

\title{Massive Protostars in a Protocluster -- A Multi-Scale ALMA View of G35.20-0.74N}

\author[0000-0001-7511-0034]{Yichen Zhang}
\affiliation{Department of Astronomy, University of Virginia, Charlottesville, VA 22904-4325, USA}
\affiliation{Star and Planet Formation Laboratory, RIKEN Cluster for Pioneering Research, Wako, Saitama 351-0198, Japan}
\correspondingauthor{Yichen Zhang}\email{yczhang.astro@gmail.com}

\author[0000-0002-6907-0926]{Kei E. I. Tanaka}
\affiliation{Center for Astrophysics and Space Astronomy, University of Colorado Boulder, Boulder, CO 80309, USA}
\affiliation{ALMA Project, National Astronomical Observatory of Japan, Mitaka, Tokyo 181-8588, Japan}

\author[0000-0002-3389-9142]{Jonathan C. Tan}
\affiliation{Department of Space, Earth \& Environment, Chalmers University of Technology, SE-412 96 Gothenburg, Sweden}
\affiliation{Department of Astronomy, University of Virginia, Charlottesville, VA 22904-4325, USA}

\author[0000-0001-8227-2816]{Yao-Lun Yang}
\affiliation{Star and Planet Formation Laboratory, RIKEN Cluster for Pioneering Research, Wako, Saitama 351-0198, Japan}
\affiliation{Department of Astronomy, University of Virginia, Charlottesville, VA 22904-4325, USA}

\author{Eva Greco}
\affiliation{Department of Astronomy, University of Virginia, Charlottesville, VA 22904-4325, USA}

\author[0000-0003-3315-5626]{Maria T. Beltr\'an} 
\affiliation{INAF -- Osservatorio Astrofisico di Arcetri, Largo E. Fermi 5, 50125 Firenze, Italy}

\author[0000-0002-3297-4497]{Nami Sakai}
\affiliation{Star and Planet Formation Laboratory, RIKEN Cluster for Pioneering Research, Wako, Saitama 351-0198, Japan}

\author[0000-0001-7378-4430]{James M. De Buizer} 
\affiliation{SOFIA-USRA, NASA Ames Research Center, MS 232-12, Moffett Field, CA 94035, USA}

\author[0000-0001-8596-1756]{Viviana Rosero}
\affiliation{National Radio Astronomy Observatory, 1003 Lopezville Rd., Socorro, NM 87801, USA}

\author[0000-0003-4040-4934]{Rub\'en Fedriani}
\affiliation{Department of Space, Earth \& Environment, Chalmers University of Technology, SE-412 96 Gothenburg, Sweden}

\author{Guido Garay}
\affiliation{Departamento de Astronom\'ia, Universidad de Chile, Casilla 36-D, Santiago, Chile}

\begin{abstract}
We present a detailed study of the massive star-forming region G35.2-0.74N with ALMA 1.3~mm multi-configuration 
observations. At 0.2$\arcsec$ (440 au) resolution, the continuum emission reveals several dense cores along a filamentary structure, 
consistent with previous ALMA 0.85~mm observations.  
At 0.03$\arcsec$ (66 au) resolution, we detect {\color{black} 22} compact sources, most of which are associated with the filament. 
Four of the sources are associated with compact centimeter continuum emission, and two of these are associated with 
H30$\alpha$ recombination line emission. 
The H30$\alpha$ line kinematics show ordered motion of the ionized gas, consistent with disk rotation and/or outflow expansion. 
We construct models of photoionized regions to simultaneously fit the multi-wavelength free-free fluxes 
and the H30$\alpha$ total fluxes. 
The derived properties suggest the presence of at least three massive young stars 
with nascent hypercompact {\Hii} regions. 
Two of these ionized regions are surrounded by a large rotating structure that feeds
two individual disks, revealed by dense gas tracers, 
such as SO$_2$, H$_2$CO, and CH$_3$OH.
In particular, the SO$_2$ emission highlights two spiral structures in one of the disks and probes the faster-rotating inner disks.
The $^{12}$CO emission from the general region reveals a complex outflow structure, 
with at least four outflows identified. 
The remaining {\color{black} 18} compact sources are expected to be associated with {\color{black} lower-mass} protostars forming 
in the vicinity of the massive stars.
We find potential evidence for disk disruption due to dynamical interactions 
in the inner region of this protocluster. The spatial distribution of the sources suggests 
a smooth overall radial density gradient without subclustering, but with tentative evidence of primordial mass segregation.
\end{abstract}


\section{Introduction}
\label{sec:intro}

Massive stars dominate the radiative, mechanical, and chemical feedback to the interstellar medium, 
by which they regulate the evolution of galaxies. Their feedback also affects nearby low-mass forming 
stars and their protoplanetary disks. In spite of their importance, many aspects of massive star formation 
remain poorly understood (see, e.g., \citealt[]{Tan14,Motte18}).

One of the challenges to form a massive star is its strong feedback on its natal gas core. In particular, 
protostellar extreme-ultraviolet (EUV) radiation can ionize the accretion disk and infall envelope to produce 
photoevaporative outflows driven by thermal pressure of the $\sim 10^4$ K ionized gas (\citealt[]{Hollenbach94}).
Detection of an {\Hii} region via free-free continuum emission and/or hydrogen recombination lines (HRLs)
has traditionally been considered as an indication of the end of the accretion phase (e.g., \citealt[]{Churchwell02}).
However, theoretical calculations have suggested that such feedback may not be strong enough to stop accretion
(e.g., \citealt[]{Keto02,Peters10,Tanaka17,Kuiper18}),
and recent high-resolution observations have revealed cases
of on-going accretion after the onset of photoionization
(e.g., \citealt[]{Maud19,Zhang19,Guzman20,Moscadelli21}).
Understanding the accretion status after the onset of photoionization is therefore crucial to constrain 
the growing efficiency under feedback during the later stages of the birth of massive stars.

Another important question is how accretion proceeds in massive star formation. 
There are a growing number of examples of accretion disks around massive protostars 
(e.g., \citealt[]{Sanchez13,Johnston15,Ilee16,Ginsburg18,Maud18,Zhang19a,Tanaka20};
see also \citealt[]{Beltran16b} and references therein),
suggesting that disk-mediated accretion is common during massive star formation. 
On the other hand, there are also some examples of massive young stars in which the accretion 
appears to occur without mediation from a 
{\color{black} large, stable disk}, but rather from multiple, more chaotic flows 
(e.g., \citealt[]{Goddi20}). 
Disks around massive protostars may be prone to gravitational instability and thus relatively likely to
fragment and form binaries or close multiple systems.
However, direct observations of sub-structures in massive protostellar disks are rare 
(e.g., \citealt[]{Motogi19,Johnston20}). Massive protostellar binary systems with a few hundreds of au separations 
have been speculated to be the result of such disk fragmentation, with continued accretion mediated by a circumbinary  
accretion disk or pseudo-disk (e.g., \citealt[]{Beltran16,Kraus17,Zhang19b}), 
although the morphology of such a circumbinary accretion structure may have evolved significantly after 
binary formation (e.g., \citealt[]{Tanaka20}). Therefore, detailed, multi-scale observations of the hierarchical structures 
of the accretion flows around massive protostars are important to understand the global accretion process 
involved in massive star formation.

On larger scales ($\gtrsim 0.01$ pc), the fragmentation of the molecular clump leads to the formation of a cluster with 
both low- and high-mass stars (see, e.g., \citealt[]{Motte18}).
The degree of fragmentation is found to be diverse among different massive star-forming regions
(e.g., \citealt[]{Palau13,Beuther18}),
which is expected to be affected by various factors such as
the initial density distribution, magnetic field strength,
degree of turbulence, and radiative heating of the cloud. 
The distributions of the massive stars within the cluster and the spatial separations of the members 
can provide important information on the fragmentation mechanism and 
to help distinguish if massive stars form via core accretion (e.g., \citealt[]{MT03}) or 
competitive accretion (e.g., \citealt[]{Bonnell01,Wang10}).
For example, \citet[]{Bonnell98} 
concluded massive stars formed preferentially in the center of the Orion Nebular Cluster (ONC).
On the other hand, \citet[]{Moser20} did not find strong evidence for primordial mass segregation in the 
massive protostar population of the massive Infrared Dark Cloud (IRDC) G028.37+00.07.
Gravitational interactions and/or radiative and mechanical feedback from massive forming stars in 
crowded protocluster environments may also affect neighboring low-mass protostars and their disks. 
It is therefore important to perform detailed studies of massive star formation in the context of star cluster formation.

G35.20-0.74N  (a.k.a. IRAS 18566+0136; hereafter G35.2)
is an ideal target to study all these processes to understand the formation
of massive stars. It is a well known massive star-forming region at a distance of 2.2 kpc (\citealt[]{Zhang09};
\citealt[]{Wu14}). Previous observations of the Atacama Large Millimeter/Submillimeter Array (ALMA)
and the Submillimeter Array (SMA) have revealed a filamentary structure
with a string of embedded cores (\citealt[]{Sanchez13,Sanchez14,Qiu13}).
CO observations have revealed a wide
outflow structure extending to $>1\arcmin$ from the central source (\citealt[]{Gibb03}; \citealt[]{Birks06})
in the direction of northeast-southwest, which is near perpendicular 
to the filament. Consistent with this wide CO outflow, near-infrared (NIR) H$_2$ observations also show
a wide outflow structure in the northeast-southwest direction, 
which may contain two outflows (\citealt[]{Caratti15}).
On the other hand, centimeter radio observations of the Very Large Array (VLA)
have revealed a collimated bipolar ionized jet
along the north-south direction (\citealt[]{Heaton88,Gibb03,Beltran16}), 
with its driving source being one of the sub-mm cores of
the filament (core B, \citealt[]{Sanchez13,Sanchez14,Beltran16}).
\citet[]{Fedriani19} revealed NIR emission spatially coincident with the radio jet in the central region.
This N-S outflow is also prominent in mid-infrared (MIR; \citealt[]{Debuizer06}).  
At these wavelengths, the emission is elongated in the N-S direction but peaked
to the north of the identified radio source and continuum core
and is thought to trace the blue-shifted outflow cavity. 
At longer wavelengths of $30-40~\mu$m, the
SOFIA-FORCAST observations have revealed the southern, far-facing
outflow cavity (\citealt[]{Zhang13,Debuizer17}).
The driving source of the N-S outflow (core B)
is reported to have a Keplerian accretion disk (\citealt[]{Sanchez13}),
with the dynamical mass estimated to be about $18\:M_\odot$.
\citet[]{Beltran16} further identified a binary system in core B (VLA sources 8a and 8b). 
By fitting the infrared spectral energy distribution (SED) assuming a single protostar, 
core B is estimated to be a protostar with a mass of 
$12-24\:M_\odot$ and a total luminosity of $(4-8)\times 10^4\:L_\odot$ (\citealt[]{Zhang13,Debuizer17,ZT18}).
In this paper, we present a detailed study of G35.20-0.74N using multi-wavelength continuum,
molecular line and hydrogen recombination line data obtained
by ALMA and VLA.

\section{Observations}
\label{sec:obs}

The observations were carried out with ALMA in Band 6 (1.3~mm)
on April 24, 2016 with the C36-3 (hereafter C3) configuration, 
on September 8, 2016 with the C36-6 (hereafter C6) configuration, 
and on November 4, 2017 with the C43-9 (hereafter C9) configuration
(ALMA project ID: 2015.1.01454.S and 2017.1.00181.S).  
The total integration times are 3.5, 6.5, and 16.8 min in the three
configurations. 
Forty-one antennas were used with baselines ranging from
15~m to 463~m in the C3 configuration, 
38 antennas were used with
baselines ranging from 15~m to 2.5~km in the C6 configuration, 
and 49 antennas were used with baselines ranging from 113~m to 13.9~km in the
C9 configuration.  
J1751+0939, J1924-2914, and J2000-1748 were used for bandpass calibration
and flux calibration.
J1830+0619 and J1851+0035 were used as phase calibrators.  
The science target was observed with single pointings, and the
primary beam size (half power beam width) was $22.9\arcsec$.

The data were calibrated and imaged in CASA (\citealt{McMullin07}).  
After pipeline calibration, we performed self-calibration
using the continuum data obtained from a 2~GHz-wide spectral window
with line-free bandwidth of 1.2~GHz. 
We first performed two phase-only self-calibration iterations with
solution intervals of 30s and 6s, and then
one iteration of amplitude self-calibration with the solution interval
equal to the scan intervals {\color{black} (ranging from $\sim1$ to $\sim5$ min)}.
We applied the self-calibration phase and amplitude solutions
to the other spectral windows.
Such self-calibration was performed
for the data of three configurations separately before any combining was performed. 
The CASA {\it tclean} task
was used to image the data, using Briggs weighting with the robust
parameter set to 0.5. 
In order to better show structures on different scales, 
below we present images made of different configuration combinations.
The resolution of the C3$+$C6 configuration is about $0\farcs25$,
and the resolution of the image of just the C9 configuration is about $0\farcs03$,
the highest achieved so far for this region at mm wavelengths.
The detailed synthesized beam sizes of images of different configuration combinations,
{\color{black} as well as their rms noise levels, are listed in Table \ref{tab:contimage}.}

{\color{black}
We also utilize the ALMA Band 7 (0.85~mm) continuum data
published by \citet[]{Sanchez13}.
Please refer to that paper for details of the observation.
Instead of directly using the published 0.85~mm continuum image,
we reproduced the continuum image using the calibrated data obtained from the ALMA
archive. From the four 2GHz-wide spectral windows, we imaged the continuum
emission using line-free channels which span a total of 0.86~GHz.
We performed two phase-only self-calibration iterations with
solution intervals of 30s and 6s, and then
one iteration of amplitude self-calibration with the solution interval
equal to the scan intervals (ranging from $\sim1$ to $\sim5$ min).
The CASA {\it tclean} task was used to image the data, using 
Briggs weighting with the robust parameter set to 0.5.

In addition, we also utilize the VLA K band (1.3 cm) and Q band (7 mm) continuum images
published by \citealt[]{Beltran16}). See that paper for details of the VLA observations.
The synthesized beams and noise levels of these images are also listed
in Table \ref{tab:contimage}.}

\section{Continuum Emission}
\label{sec:continuum}

\subsection{Lower-resolution 1.3~mm Continuum}
\label{sec:continuum_lowres}

\begin{figure}
\begin{center}
\includegraphics[width=1\textwidth]{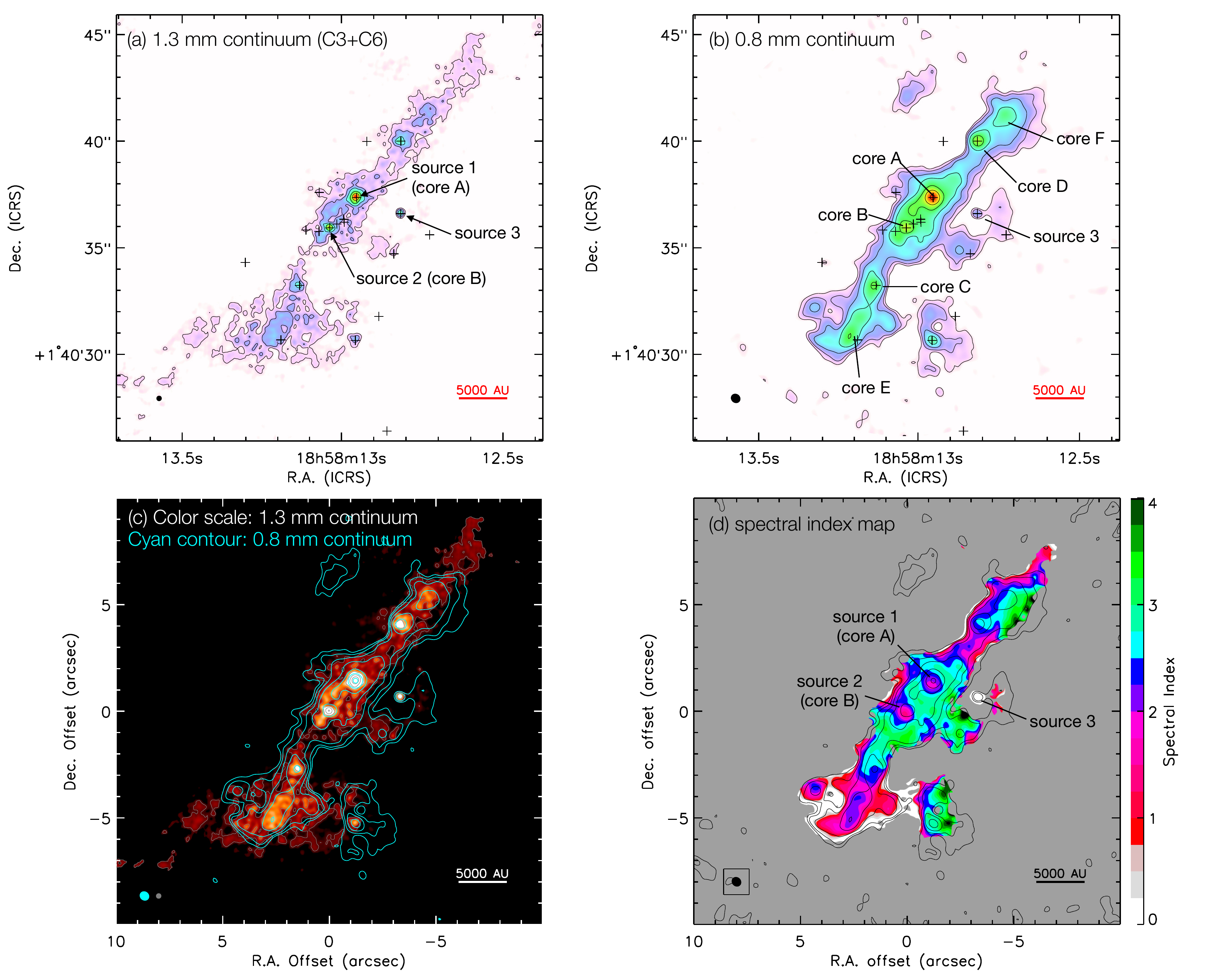}\\
\caption{\color{black} 
{\bf (a)}: Lower resolution ALMA 1.3~mm (Band 6) continuum image of the
G35.2 star forming region, obtained by combining data from C3 and C6 configurations.
The contour levels are $5\sigma\times2^n$ ($n=0,1,...$), 
with $1\sigma=0.17~\K$ ($0.48~\mJybeam$) 
the synthesized beam is $0\farcs254\times 0\farcs250$ ($\pa=-36.3^\circ$).
The crosses mark the position of the compact sources identified from
high-resolution 1.3~mm continuum image and the names of the three main sources
are labeled (see Figure \ref{fig:contmap_config} and Table \ref{tab:cont}).
{\bf (b)}: The 0.85~mm continuum image re-imaged using the data published by \citet[]{Sanchez14}.
The synthesized beam is $0\farcs47\times 0\farcs42$ ($\pa=51.4^\circ$)
and contour levels of $5\sigma\times2^n$ ($n=0,1,...$) with
$1\sigma=0.039~\K$ ($0.74~\mJybeam$).
The cores A$-$F identified from the 0.85~mm image by \citet[]{Sanchez14} are
labeled. The crosses are the same as panel a.
{\bf (c)}: Comparison of the ALMA 1.3~mm continuum image (color scale and grey contours) 
and the 0.85~mm continuum image (cyan contours).
{\bf (d)}: Spectral index map between 1.3 and 0.85~mm shown in color scale,
overlaid with the 1.3~mm continuum image with the same contour levels as in panel a.
The spectral index is defined as $\alpha_\nu=\log(I_{\nu_1}/I_{\nu_2})/\log(\nu_1/\nu_2)$,
where $\nu_1=234~\mathrm{GHz}$ and $\nu_2=343~\mathrm{GHz}$.
Only the regions with $>10\sigma$ continuum emissions and spectral index errors $<0.5$
are shown in the figure.
The origin of the R.A. and Dec. offsets in panels c, d are set to be core B
($\alpha_\mathrm{ICRS}=18\h58\m13\s.037$,
$\delta_\mathrm{ICRS}=+1^\circ40\arcmin35\farcs931$).}
\label{fig:contmap}
\end{center}
\end{figure}

{\color{black} Panels a and b of Figure \ref{fig:contmap} show the ALMA 1.3~mm continuum image, 
combining the data of C3 and C6 configurations, and the reproduced 0.85~mm continuum image.
In general, the 1.3~mm continuum morphology is consistent with that of the 0.85~mm continuum, 
i.e., showing multiple cores along the filamentary structure. 
Compared to that presented by \citealt[]{Sanchez13}, the new 0.85~mm continuum image
shows a compact source to the west of core A (labeled as source 3 in the figure; 
see \S\ref{sec:continuum_highres}). This source is seen as a compact continuum source 
in the 1.3~mm continuum image, as well as all the VLA images, but
was not seen in the previously presented 0.85~mm image.
Detection of this source in the reproduced 0.85~mm continuum image is probably due to the improved sensitivity
brought by the self-calibration.
The new 0.85~mm image also shows more diffuse emission in the southern end of the filament, which
is also seen in the 1.3~mm continuum image.
Panel c shows a comparison between the 1.3 and 0.85~mm continuum. 
The overall morphology and the locations of the bright sources/regions
coincide very well in the two images.}

{\color{black} Figure \ref{fig:contmap}d shows the spectral index map between 
ALMA 1.3 and 0.85~mm continuum. 
Here the spectral index is defined as $\alpha_\nu=\log(I_{\nu_1}/I_{\nu_2})/\log(\nu_1/\nu_2)$, 
where $I_{\nu}$ is the intensity at the frequency $\nu$, 
$\nu_1=343$~GHz (0.85~mm) and $\nu_2=234$~GHz (1.3~mm). 
The spectral index map is obtained with the 0.85~mm data and 1.3~mm C3+C6 data,
using multi-frequency synthesis (mtmfs) in the CASA {\it tclean} task with 
$nterms = 2$. 
In order to derive the spectral index map more accurately, 
we only used the 1.3~mm data with baselines between 20~k$\lambda$ and 440~k$\lambda$
to match the $uv$ range of the 0.85~mm data.}

Typical dust continuum emission has spectral indices of $\alpha_\nu>2$, 
with $\alpha_\nu=2$ in the optically thick case and $3\lesssim\alpha_\nu\lesssim4$ 
in the optically thin case (assuming a typical dust emissivity spectral index of $1-2$). 
Free-free continuum emission has spectral indices of $-0.1 \lesssim \alpha_{\nu}<2$, 
with $\alpha_\nu=2$ in the optically thick case and $\alpha_\nu\approx -0.1$ in the optically thin case.
Cores A and B have spectral indices below 2, which suggests a contribution of free-free emission,
expected for forming massive stars.
However, the observed spectral indices are not far from 2.
Optically thick dust emission or a combination of free-free and dust emission can also lead to such 
spectral indices. The regions with spectral indices above 2, which are dominated by dust emission, 
are mostly around cores A and B along the filament, suggesting dusty envelopes around the small 
{\Hii} regions created by the forming massive stars.
{\color{black} Cores C, D and F all have spectral indices close to 3, dominated by dust emissions, while 
core E has a spectral index slightly above 2, indicating a higher free-free fraction or high optical depth.
Source 3 has a very flat SED between 1.3 and 0.85~mm, suggesting it
is dominated by free-free emission.}

\subsection{Long-baseline 1.3~mm Continuum}
\label{sec:continuum_highres}

\begin{figure}
\begin{center}
\includegraphics[width=1\textwidth]{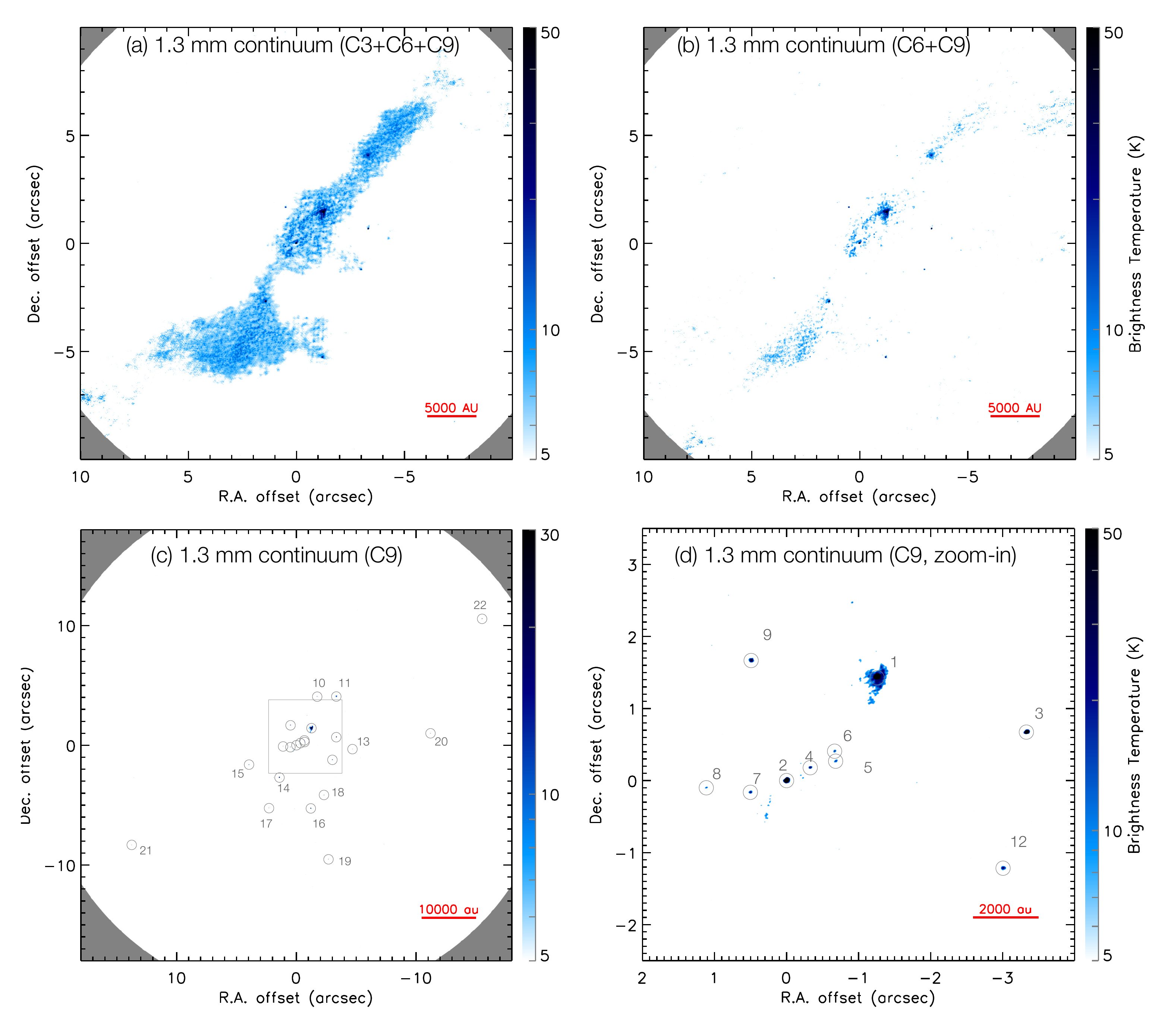}\\
\caption{High resolution ALMA 1.3~mm continuum images from different configurations.
{\bf (a)}: Combined image from the C3, C6 and C9 configurations.
{\bf (b)}: Combined image from C6 and C9 configurations.
{\bf (c)}: Image from C9 configuration alone.
{\bf (d)}: Same as panel c, but zoomed in to the central region.
{\color{black} The synthesized beam size of each image is listed in Table \ref{tab:contimage}.}
{\color{black} The color scale (same for all the panels) is adjusted to emphasize the compact continuum sources.
In panels a and b, only the regions with primary beam response of $>0.5$ are shown.
In panels c, all the regions with primary beam response of $>0.1$ are shown.}
The identified compact continuum sources are marked with circles and labeled
with numbers in panels c and d.
}
\label{fig:contmap_config}
\end{center}
\end{figure}

\begin{figure}
\begin{center}
\includegraphics[width=0.5\textwidth]{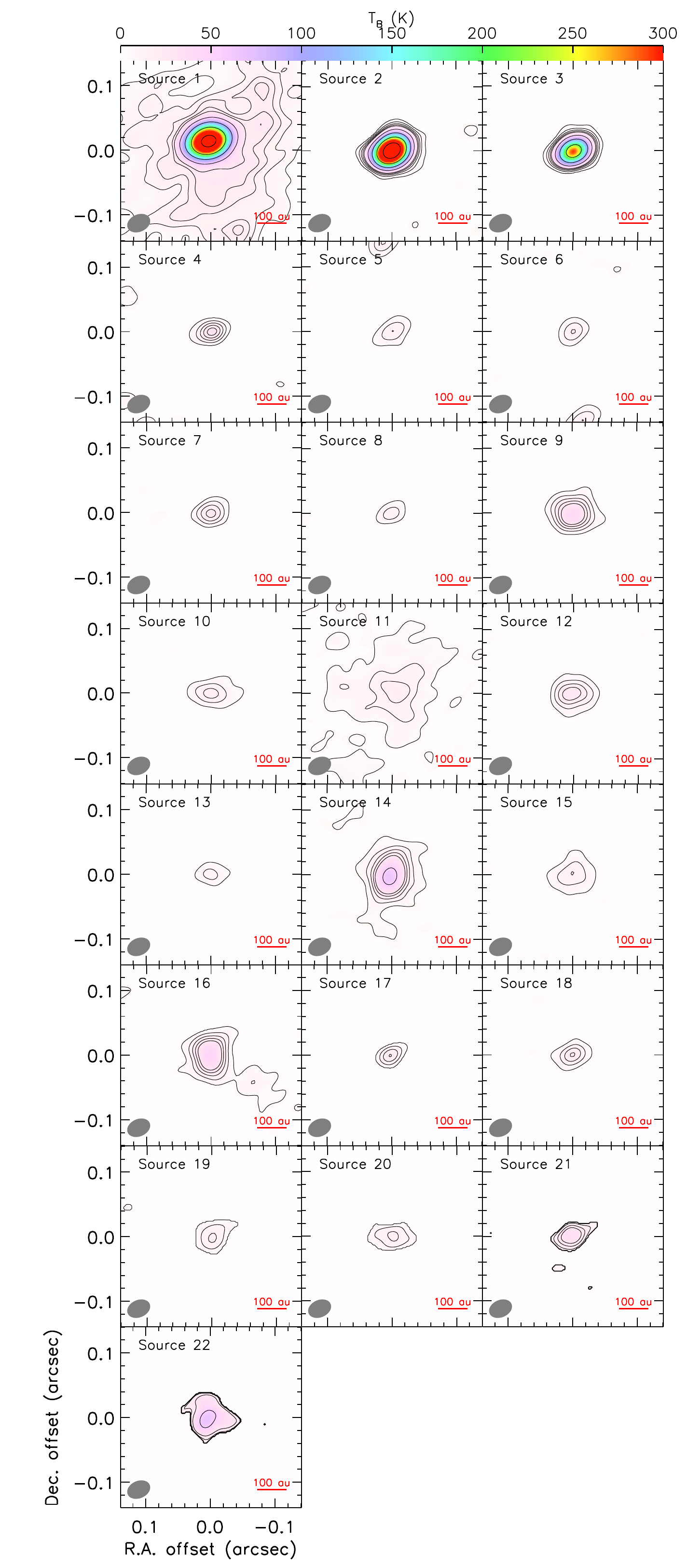}\\
\caption{ALMA long-baseline 1.3~mm continuum images of all the compact sources identified 
(see Table \ref{tab:cont}).
The contour levels are 4, 8, 12, 16, 20, 40, 80, 160, 320 times the rms noise 
$1\sigma=0.0608~\mJybeam$ (1.37 K).
The synthesized beam is shown in the bottom-left corner of each panel.}
\label{fig:low-mass}
\end{center}
\end{figure}

Figure \ref{fig:contmap_config} shows the high-resolution view of the 1.3~mm continuum emission. 
From the long-baseline continuum data (C9 configuration), 
we are able to identify a total of {\color{black} 22} compact sources 
{\color{black} in the whole field of view with a primary beam response $>0.1$
(20 sources within the radius of primary beam response $>0.5$)}. 
The {\color{black} candidates of compact sources are first} 
identified with maximum intensities $>5\sigma_\mathrm{noise}$ 
($1\sigma_\mathrm{noise}=0.06~\mJybeam$) and sizes estimated for regions with 
intensities $>4\sigma_\mathrm{noise}$ larger than that of one beamsize.
{\color{black} We further removed the emission peaks caused by the extended filament emissions 
or the bright patches of the residual pattern, and made sure that
the remaining compact sources clearly stand out from the residual pattern fluctuation.
Note that the residual pattern close to the central region has a  fluctuation level of 
$1\sigma_\mathrm{residual}=0.1~\mJybeam$.
And the confirmed compact sources all have peaks $\gtrsim 6\sigma_\mathrm{residual}$ ($\gtrsim10\sigma_\mathrm{noise}$).} 
Figure \ref{fig:low-mass} shows the continuum image of each source. 
Table \ref{tab:cont} lists the identified compact sources, as well as their associated cores and corresponding 
VLA compact sources. The positions of these sources are also marked in 
{\color{black} panels a and b of Figure \ref{fig:contmap}} to 
show their association with the larger scale structures.

Most of the compact continuum sources are along the filament, especially around cores A and B. 
Among the six cores previously identified from the lower-resolution 0.85~mm continuum image, 
five (except core F) have associated compact continuum sources.
{\color{black} ALMA source 3, which is a bright and compact source in both the
long-baseline and lower-resolution 1.3~mm images
is now also shown in our new 0.85~mm continuum image (Figure \ref{fig:contmap}b).
ALMA source 16 is also seen to be associated with a dense core structure in the new 0.85~mm continuum,
which was not reported previously.}

\begin{figure}
\begin{center}
\includegraphics[width=0.8\textwidth]{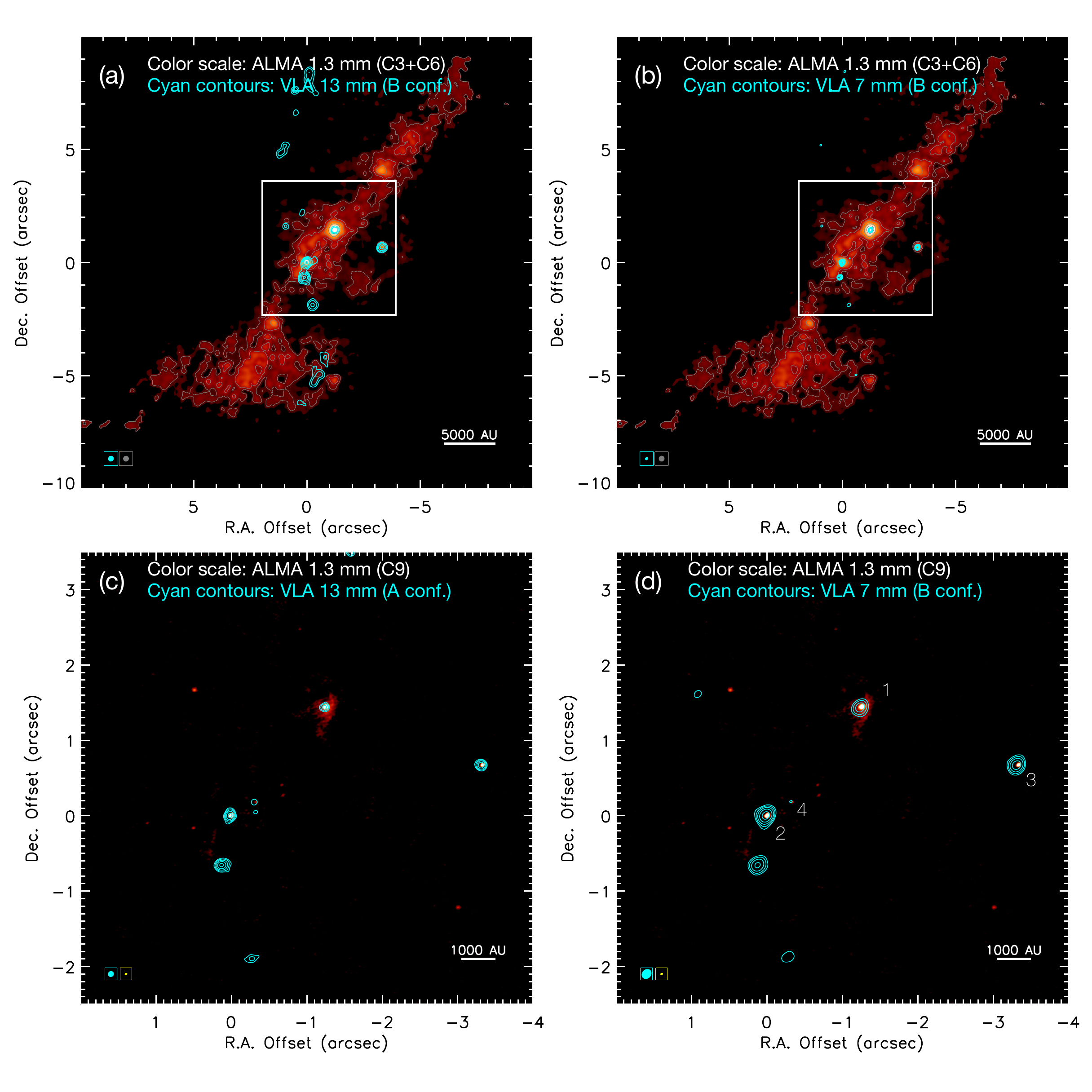}\\
\caption{ALMA 1.3~mm continuum images (color scale) overlaid
with VLA 1.3 cm (left column) and 7 mm (right column) continuum images (cyan contours). The wavelengths and configurations of the images are labeled in each panel. Panels c and d show zoom-in views of the central region
(the white squares in panels a and b).
{\color{black} The synthesized beams of the images are listed in Table \ref{tab:contimage}.
The contours start from $5\sigma$ and have intervals of $10\sigma$. The $1\sigma$ rms noise levels of each image
is listed in Table \ref{tab:contimage}.}
The compact continuum sources identified in ALMA 1.3~mm images with 
corresponding VLA emission are labeled in panel d.
}
\label{fig:contmap_VLA}
\end{center}
\end{figure}

Figure \ref{fig:contmap_VLA} overlays the VLA 1.3 and 7~mm continuum emission images 
in the B and A configurations with the 1.3~mm low and high resolution continuum images. 
As previously pointed out, the large-scale VLA continuum emission mostly traces a radio jet originating 
from core B (source 2 in our list). As seen in this figure, ALMA sources 1, 2, 3, and 4 have associated 
emission in the VLA bands. ALMA sources 2 and 4, which are associated with VLA sources 8a and 8b, respectively, 
are likely members of a binary system, as suggested by \citet[]{Beltran16}.

ALMA sources 1$-$4 are the only four 1.3~mm compact sources associated with VLA compact emission 
(VLA source 3, 8a, 1, and 8b). 
The other emission knots identified in the VLA continuum maps do not have any counterpart compact source at 1.3~mm. 
These sources mostly have flat or negative spectral indices in VLA bands (see \citealt[]{Beltran16}). 
Extrapolating from the VLA bands to ALMA 1.3~mm, their fluxes should be below the continuum detection limits of 
ALMA unless there are additional dust contributions. Therefore, the non-detection of these VLA sources in ALMA 1.3~mm 
supports the idea that they are jet knots caused by shock ionization and are not protostars (\citealt[]{Beltran16}). 
Such a scenario is also supported by the fact that \citet[]{Fedriani19} detected atomic ([Fe II] and Br$\gamma$) and 
molecular (H$_2$) emission associated with shocks towards the north side of core B, 
i.e., the blue-shifted outflow (also see \S\ref{sec:outflow}).
{\color{black} On the other hand, the other ALMA compact sources that are not detected in VLA bands 
(ALMA sources 5$-$22) should be lower-mass protostars 
(or relatively massive but young sources which have not yet reached main sequence to produce detectable photoionized regions) 
forming along with the photoionizing massive stars.}

\begin{figure}
\begin{center}
\includegraphics[width=\textwidth]{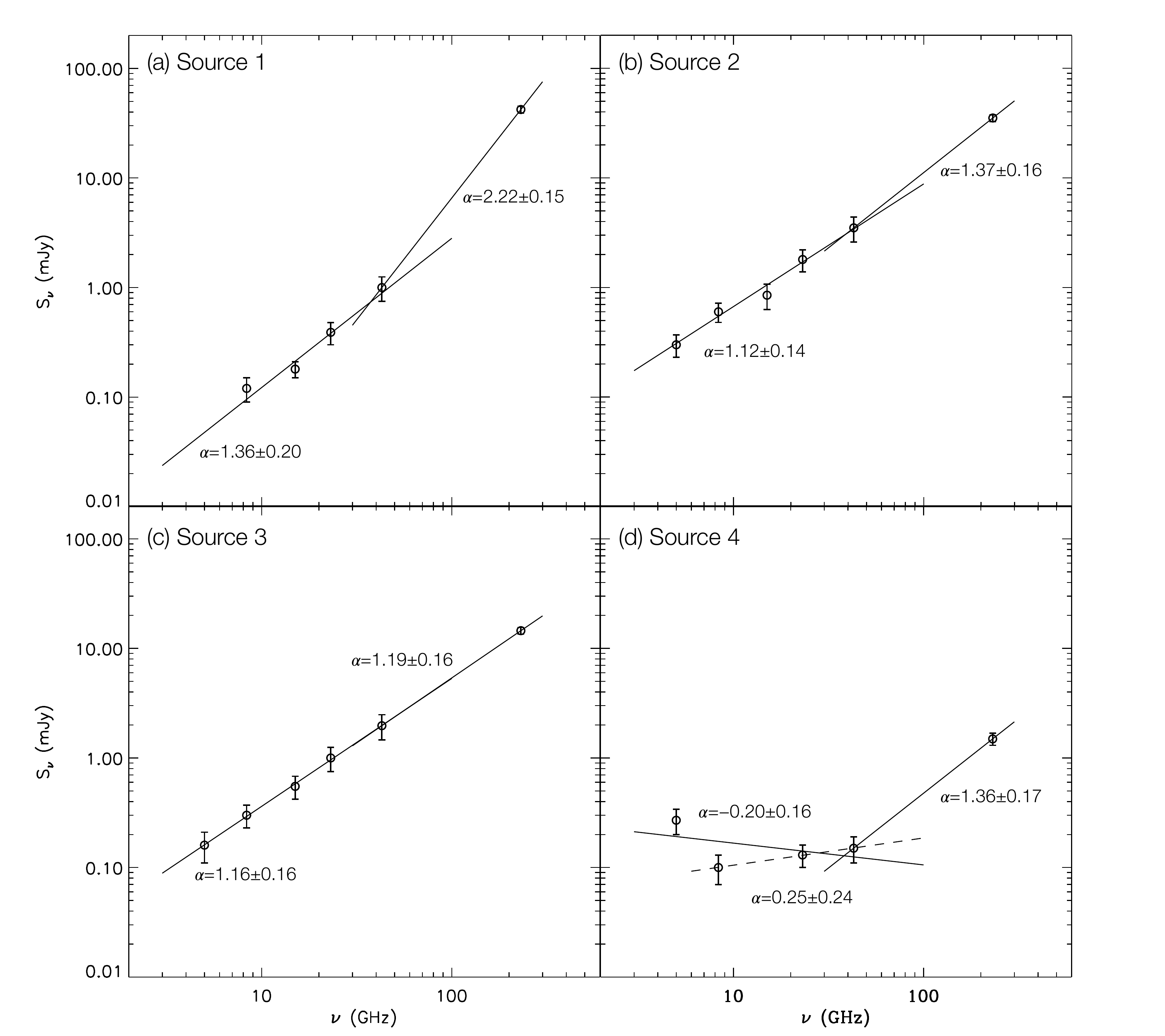}\\
\caption{Spectral energy distributions (symbols and error bars) 
comprising of VLA and ALMA 1.3~mm continuum flux densities,
for the 1.3~mm compact sources with VLA detections, i.e. ALMA sources 1$-$4.
The ALMA 1.3~mm flux densities are measured from C9 configuration image and are listed
in Table \ref{tab:cont}. The VLA flux densities are taken from \citet[]{Beltran16}.
Additional 7\% of uncertainties due to absolute flux calibration are added to the ALMA 1.3~mm flux densities.
The power-law fit to the VLA fluxes, and between VLA 7~mm and ALMA 1.3~mm, are shown by
the solid lines.
In source 4 (panel d), the dashed line show the power-law fit to the VLA fluxes excluding the 6~cm flux.
}
\label{fig:sed}
\end{center}
\end{figure}

For the ALMA sources with corresponding cm emission (sources 1$-$4), 
Figure \ref{fig:sed} shows their continuum SEDs from 6~cm to 1.3~mm.
The VLA flux densities are obtained from \citet[]{Beltran16}.
{\color{black} The 1.3~mm flux densities are obtained from the long-baseline (C9) data.}
These sources have spectral indices from 6 cm to 7 mm (VLA bands) either
near flat (source 4) or slightly higher than 1 (sources 1$-$3), consistent with free-free emission.
Free-free emission tends to have its spectral index decrease as the frequency becomes higher.
However, in all these sources, the spectral indices tend to increase at 1.3 mm (especially in sources 1 and 4), 
indicating dust contributions.
{\color{black} As the free-free emission in these sources are highly compact, the different baseline ranges in
these data have minimum effects on the SED slope. For the 1.3~mm data, we also only focus
on the most compact emission structures here, so filtering-out of the extended dust emission 
in the C9 configuration data does not affect our discussion.}
The compact dust emission confirms that they are forming stars rather than jet knots.
For sources 1$-$3, the existence of bright free-free and dust emission is consistent with 
forming massive stars creating small ionized regions via photoionization. 
This was speculated by \citet[]{Beltran16} and is now further supported by the ALMA observations.
{\color{black} For source 4, while such a scenario is also possible, its 1.3 mm flux is on similar level as 
the other low-mass sources in the region.
If source 4 is not yet massive enough to produce enough ionizing photons, 
the free-free emission could be caused by shock ionization of an unresolved jet,
which is seen in protostellar sources with a wide range of masses (e.g. \citealt[]{Tychoniec18,Anglada18}).}
We will discuss these SEDs in detail in \S\ref{sec:spec}.

\section{Hydrogen Recombination Lines}
\label{sec:HRL}

H30$\alpha$ hydrogen recombination line emission is detected toward
sources 2 and 3. Figure \ref{fig:HRL} shows the integrated H30$\alpha$ 
emission maps of these two sources and their H30$\alpha$ spectra at the continuum peak positions. 
The H30$\alpha$ peaks coincide well with the continuum peaks, which supports the existence of small 
ionized regions at the centers of sources 2 and 3 that contribute both to the H30$\alpha$ emission and 
the free-free continuum emission. 
{\color{black} Even though the H30$\alpha$ emission in source 2 is only detected with low S/N ($6\sigma$), the fact that
the broad spectral feature is spatially coincident with the continuum emission suggests that
the H30$\alpha$ detection is likely real.}
The H30$\alpha$ lines are relatively well fitted with Gaussian profiles. 
The H30$\alpha$ central velocities are derived to be $29.4\pm2.2~\kms$ and $30.2\pm0.2~\kms$ for 
sources 2 and 3, respectively, which are similar to each other within
the uncertainties, and also consistent with the systemic velocity of source 2 ($30.0\pm0.3~\kms$) estimated by fitting the 
Keplerian disk kinematics traced in different molecular lines (\citealt[]{Sanchez13}). The line widths are derived to be 
$86.1\pm5.2~\kms$ and $45.0\pm0.6~\kms$ for source 2 and 3 respectively. 
The line width of $86~\kms$ is on the upper edge of observed recombination line 
widths in hypercompact {\Hii} (HC{\Hii}) regions (e.g. \citealt[]{Hoare07}). 
Although, larger line widths have been reported in some HC{\Hii} regions 
(e.g. NGC 7538 IRS 1; \citealt[]{Sewilo04}); 
a line width of $\lesssim50~\kms$ is more typical for such regions.

\begin{figure}
\begin{center}
\includegraphics[width=\textwidth]{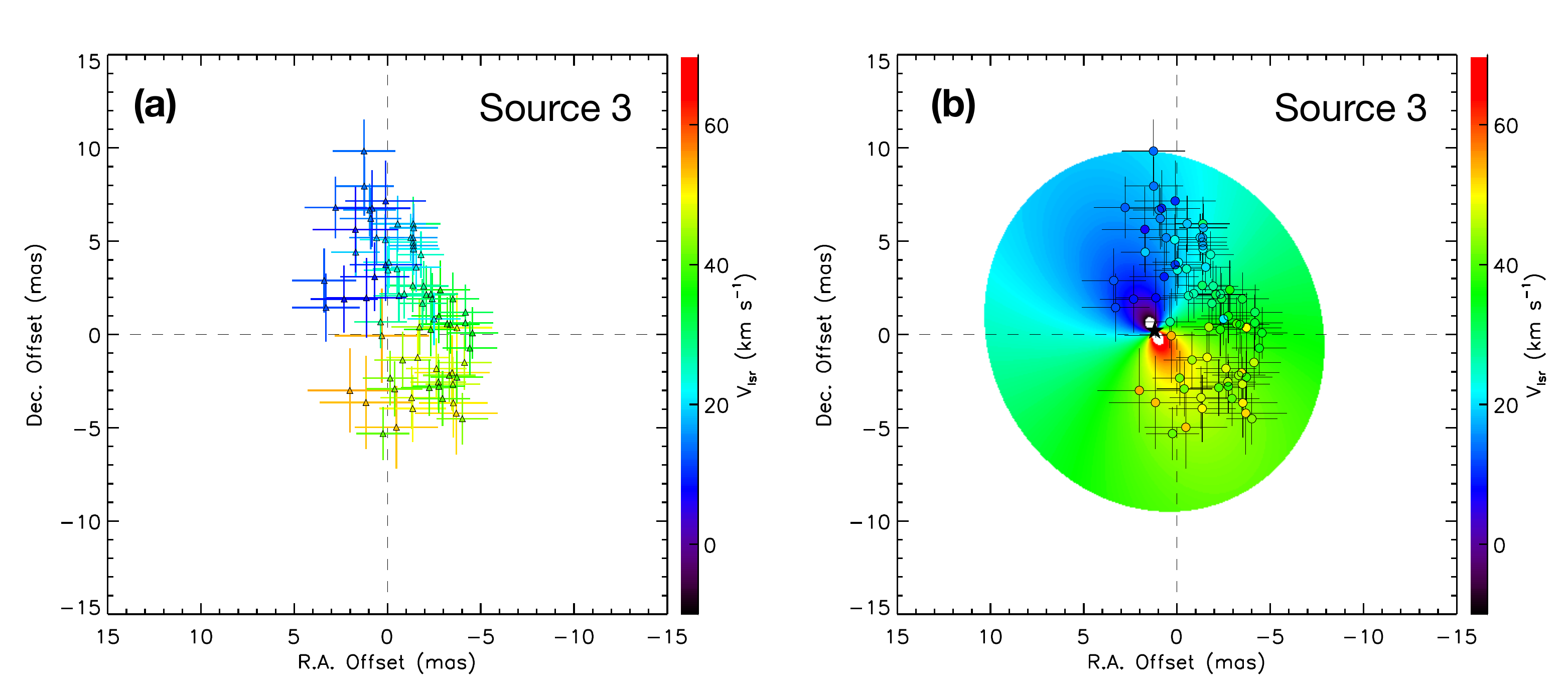}\\
\caption{{\bf (a)}: Distribution of the H30$\alpha$ emission centroids (triangles with error bars)
in source 3. 
Only channels with peak intensities $>10\sigma$ ($1\sigma=1.8~\mJybeam$) are included. 
The position offsets are relative to the continuum peak. 
The line-of-sight velocities are shown by the color scale.
{\bf (b):} The best-fit disk model for the H30$\alpha$ centroid distribution in source 3
(see the text for more details).
}
\label{fig:HRL_vel}
\end{center}
\end{figure}

Panels e and f of Figure \ref{fig:HRL} show the velocity structures of the H30$\alpha$ lines in 
sources 2 and 3. The integrated maps of the blue- and red-shifted emission
show shift of emission with velocity in these two sources.
For source 3, the high S/N further allows us to determine the shift of the 
emission centroid with channel velocity (Figure \ref{fig:HRL_vel}). 
The centroid positions are determined by fitting Gaussian ellipses to the H30$\alpha$ emission 
in the C3+C6+C9 configurations at channels with peak intensities $>10\sigma$. 
Following the method used by \citet[]{Zhang19}, the uncertainties of the centroid positions are 
determined from the data S/N as well as the phase noise in the passband calibrator. 
The centroid distributions and kinematics are consistent with a rotating structure with a radius of about 
5 mas (11 au). The elliptical distribution of the centroids, and the fact that the centroids with the highest 
velocities are closer to the center, are similar to what have been seen in molecular lines around 
massive protostars (e.g., \citealt[]{Sanchez13,Ilee16}), which indicates faster rotation velocities
near the central source, as expected for a Keplerian disk.

Following the methods used by \citet[]{Sanchez13} and \citet[]{Ilee16}, we fitted the H30$\alpha$ centroid distribution 
with a model of a Keplerian disk. 
The model has six free parameters: mass of the central object, the right ascension and declination offsets of the center, 
disk position angle, disk inclination, and systemic velocity. 
The fitting minimizes the differences between the channel velocity and the line-of-sight velocity
of disk rotation at the peak position of each channel. 
However, in our case, different combinations of disk inclination and central mass can generate
fitting with similarly good qualities. Therefore we cannot well constrain these two properties solely 
by such fittings. To break the degeneracy, we set the central mass to $16~M_\odot$, which is estimated 
from the free-free continuum (see \S\ref{sec:source3}). 
Figure \ref{fig:HRL_vel}(b) shows the best-fit model. 
This model has a systemic velocity of $\vsys=+30~\kms$, 
R.A. offset of $\Delta\alpha=1.2~\mathrm{mas}$, 
Dec. offset of $\Delta\delta=0.2~\mathrm{mas}$, 
position angle of the disk major axis of $\pa=28^\circ$, 
and disk inclination of $i=26^\circ$ between the disk plane and plane of sky. 
Although such a simple model fitting cannot rule out other possibilities for the 
observed H30$\alpha$ kinematic pattern, 
it shows that disk rotation around a forming massive star can be a valid explanation.

While the velocity gradient of the ionized gas in source 3 could trace a rotating disk, 
{\color{black} it is more difficult to tell whether the velocity gradient seen in source 2 (Figure \ref{fig:HRL}e) is
real or not, or what its possible origin is if it is real, due to the low S/N.} 
The direction of the {\color{black} tentative} H30$\alpha$ velocity gradient in source 2 is at a position angle of $\pa\approx 25^\circ$. 
The disk around source 2 has a position angle of $\pa\approx -25^\circ$ (see \S\ref{sec:line}; also \citealt[]{Sanchez13}), 
and the ionized jet originating from source 2 has a position angle of $\pa\approx20^\circ$ close to the source, 
as the jet has an S-shape morphology (\citealt[]{Beltran16}).
Therefore the direction of the velocity gradient of the H30$\alpha$ line is consistent with the jet direction. 
However, the line-of-sight velocity directions seen in the H30$\alpha$ line (red-shifted in north and blue-shifted in south) 
are opposite from the outflows.
Molecular line observations have shown that the blue-shifted molecular outflow is toward the north and northeast, 
while the red-shifted outflow is toward the south and southwest (\citealt[]{Gibb03,Birks06,Qiu13}; see \S\ref{sec:outflow}). 
The atomic line observations show that the jet toward the north is blue-shifted (\citealt[]{Fedriani19}). 
The jet-associated H$_2$O masers have also been reported to be red-shifted ($\vlsr>30~\kms$) toward the south. 
Therefore, despite the fact that the velocity gradient orientation is similar to the jet direction, it is difficult to confirm that 
the velocity gradient is mainly tracing outflow motion. 
One possibility is that the observed H30$\alpha$ velocity gradient is affected by the disk's rotation (red-shifted on the northern side).
In fact, in this source, the Br$\gamma$, an NIR hydrogen recombination line, does show one high velocity component associated
with the jet and a slow component associated with the disk (\citealt[]{Fedriani19}),
supporting that the ionized gas in this source is in both outflow and disk.
Another possibility is that the jet is close to edge-on and therefore the inclination of the innermost part of the outflow 
is different from that of the outflows at larger distances.

As there are more and more 100-au-scale observations toward massive star forming regions, 
kinematics of photoionized gas at disk scales around forming massive stars traced 
by recombination lines have started to be reported. 
This includes ionized disks (some are surrounded by neutral disks traced by molecular lines) 
and ionized outflows just launched from the disk (some are co-rotating with the disk)
(e.g., \citealt[]{Zhang17}; \citealt[]{Zhang19a,Zhang19}, \citealt[]{Guzman20,Jimenez20,Moscadelli21}). 
These observations show the effectiveness of using recombination lines as a tracer of the ionized gas 
in the innermost region around forming massive stars to study accretion and 
photoionization feedback processes in massive star formation.

\section{Fitting the Continuum and H30$\alpha$ Fluxes}
\label{sec:spec}

For sources $1-4$, Figure \ref{fig:sed} shows that most SEDs in the VLA bands have spectral indices between 0 and 2, 
indicating partially optically thick free-free emission. 
The optical depth of the free-free emission is dependent on the frequency (see Appendix \ref{sec:appA}). 
For an ionized region with a uniform electron density, a spectral index $\alpha\approx 1$ is only seen in a relatively 
narrow transition frequency range. However, in our detected sources, 
the intermediate spectral indices are seen over a wide range of frequency, 
which requires a density gradient in the ionized gas (e.g. \citealt[]{Keto03}).

We construct a simple model to explain the observed free-free continuum. 
The total H30$\alpha$ fluxes are also used to constrain the model. 
The model details are introduced in Appendix \ref{sec:appA}. 
In the model, we assume the emitting region is a circular disk on the plane of sky, and the emission measure 
($\elecm=\int n_e^2 dl$) follows a power law dependence on the radius ($\elecm\propto r^{\rho_\mathrm{EM}}$). 
The three free parameters of the model are the electron temperature of the ionized gas ($T_e$), 
the emission measure at the radius of $10~\au$ ($\elecm_{\mathrm{10au}}$), 
and the power-law index of the emission measure distribution $\rho_\mathrm{EM}$. 
The fitting results are listed in Table \ref{tab:sedmodel}. Below we discuss the fitting results of each source in detail.

\subsection{ALMA Source 2}
\label{sec:source2}

\begin{figure}
\begin{center}
\includegraphics[width=\textwidth]{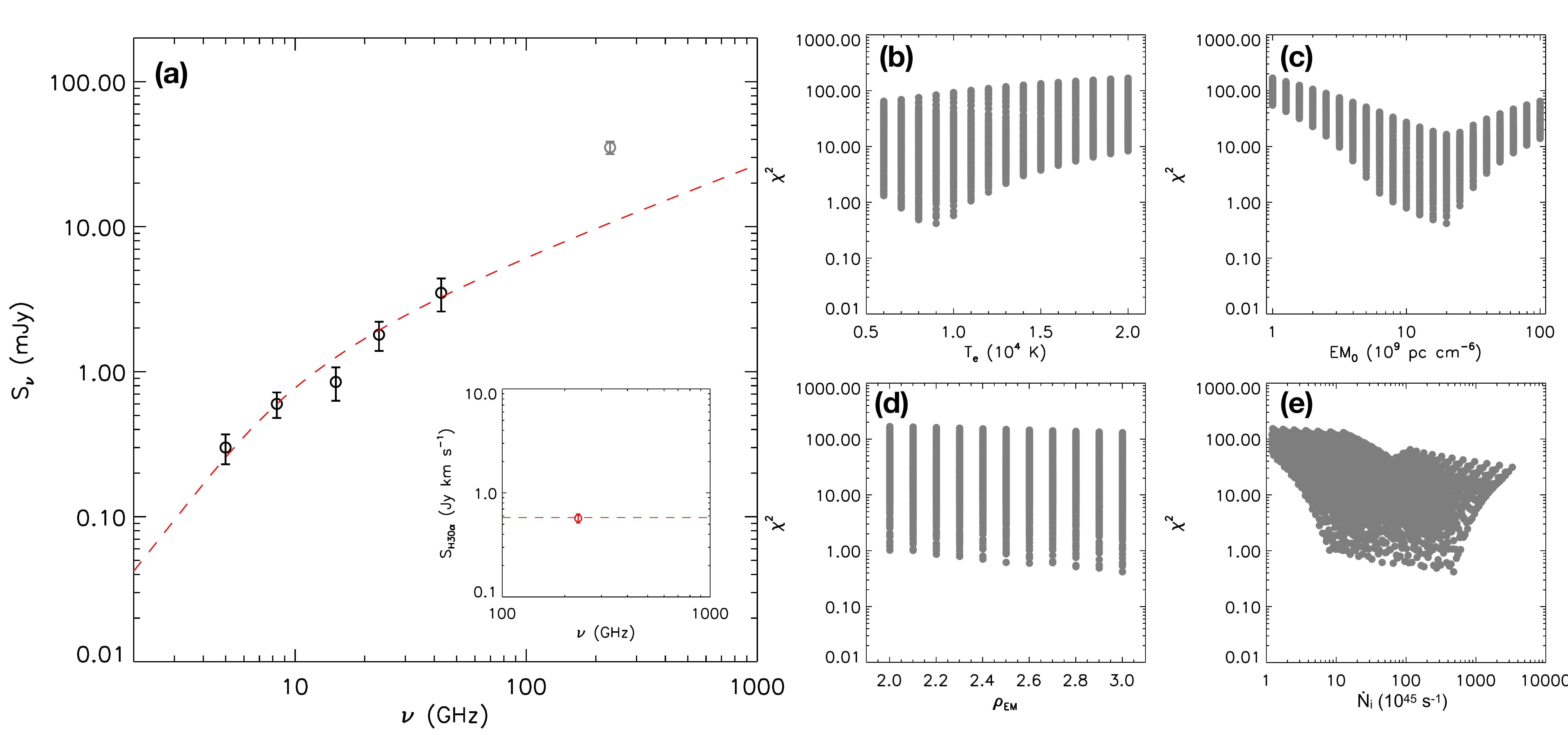}\\
\caption{{\bf (a):} The continuum spectral energy distribution 
of source 2 for the
 VLA bands and ALMA 1.3~mm continuum (data points).
 The red dashed line shows the free-free emission of the best-fit model.
 ALMA 1.3~mm continuum (shown in grey) is not used in the fitting.
 The inset panel shows the H30$\alpha$ flux integrated over velocity and area
 (red data point) and the model fit (red dashed line).
 The model fits the continuum and H30$\alpha$ fluxes simultaneously.
{\bf (b)$-$(d):} $\chisq$ distribution with model parameters: electron temperature
 $T_e$ (panel b), the emission measure $\elecm_0$ at the radius of $10~\au$
 (panel c), and the power-law index of the emission measure distribution 
 $\rho_\mathrm{EM}$ (panel d).
 {\bf (e):} $\chisq$ distribution with the ionizing photon rate $\dot{N}_\mathrm{i}$
 calculated from $T_e$, $\elecm_0$, and $\rho_\mathrm{EM}$.}
\label{fig:sed_source2}
\end{center}
\end{figure}

For source 2, we simultaneously fit the VLA continuum fluxes 
(assuming they are free-free dominated) and H30$\alpha$ flux. 
Figure \ref{fig:sed_source2} shows the best-fit model and the $\chisq$ distribution in the parameter space. 
The 1.3~mm continuum flux is not used in the fitting as it contains significant dust contribution. 
The ionized gas temperature and emission measure are both well constrained by the fitting, 
with the best-fit model having $T_e=9,000~\K$ and $\elecm_{\mathrm{10au}}=2\times 10^{10}~\pccm$. 
The power-law index of the emission measure gradient is not well constrained. 
The best-fit model has $\rho_\mathrm{EM}=3$, 
but all $\rho_\mathrm{EM}$ values between 2 and 3 can produce similarly good fits. 
{\color{black} Note that $\rho_\mathrm{EM}=2$ corresponds to a power-law index of $\rho_{n_e}=1.5$
for the electron density gradient, which is consistent with a photoionized disk,
while $\rho_\mathrm{EM}=3$ corresponds to a power-law index of $\rho_{n_e}=2$ in the electron density gradient,
which is consistent with an expansion flow (see Appendix \ref{sec:appA}).
The best-fit model with $\rho_\mathrm{EM}=3$ is therefore consistent with the wide recombination line 
(see \S\ref{sec:HRL}).} 
The derived $\elecm_{\mathrm{10au}}=2\times 10^{10}~\pccm$ also leads to an electron density of 
$n_e\simeq2.0\times 10^7(l/{\mathrm{10au}})^{-1/2}~\mathrm{cm}^{-3}$ 
at 10 au scale ($l$ is the line-of-sight scale of the ionized region).
The compact size ($< 100~\au$, seen from the image) of the ionized region, 
the derived emission measure and electron density, 
as well as the H30$\alpha$ line width of 80~$\kms$, 
are all consistent with a HC{\Hii} region (e.g., \citealt[]{Hoare07}).

From the fitted parameters $T_e$, $\elecm_{\mathrm{10au}}$, and $\rho_\mathrm{EM}$, 
we further estimate the ionizing photon rate $\dot{N}_\mathrm{i}$ (see Appendix \ref{sec:appA}).
Since $\dot{N}_\mathrm{i}$ is sensitive to $\rho_\mathrm{EM}$ (see the last term in Eq. \ref{eq:ni}) and 
$\rho_\mathrm{EM}$ is not well constrained,
the models with similarly good fit give a range of 
$\dot{N}_\mathrm{i}=(0.8-48)\times 10^{46}~\mathrm{s}^{-1}$,
with $\dot{N}_\mathrm{i}=48\times 10^{46}~\mathrm{s}^{-1}$ being the best model
(Figure \ref{fig:sed_source2}e).
For zero-age main sequence (ZAMS) stars, 
this corresponds to a stellar mass of 11.7 to $18.4\:M_\odot$ (\citealt[]{Davies11}), 
i.e., spectral type of B0.5$-$O9.5 (\citealt[]{Mottram11}). 
Our estimates of the emission measure, ionizing photon rate, and the stellar mass, 
are higher than that estimated by \citet[]{Beltran16}
($\elecm-2.7\times 10^8~\pccm$, $\dot{N}_\mathrm{i}=1\times 10^{45}~\mathrm{s}^{-1}$, and
B1 type ZAMS star).
This is because their estimation is based on the assumption of optically thin free-free emission,
and a uniform density distribution.  
In our model, we consider a gradient in the emission measure 
(and therefore optical depth of free-free emission)
which leads to higher values of ionizing photon rate and stellar mass.
We note that the derived ZAMS mass should be treated as a reference mass
{\color{black}
because a forming massive star that is still in a bloated phase before reaching the main sequence
requires more mass than a ZAMS star to produce same amount of ionizing photons
(e.g. \citealt[]{Hosokawa10}).
}
Detailed protostellar evolution calculations taking into account 
various accretion histories of massive star formation show
that accreting massive young stars with a wide range of mass
have similar levels of ionizing photon rates 
(e.g., \citealt[]{ZTH14,ZT18,Tanaka16}), 
which we will discuss in more details in \S\ref{sec:mass}.

The best-fit model gives a free-free flux of $10~\mJy$ at 1.3~mm, 
leaving about $25~\mJy$ to be the dust emission at 1.3~mm. 
Such a 1.3~mm dust emission flux leads to about $0.9~\mJy$ of dust emission at 7 mm
assuming an optically thick dust spectral index of 2. 
The dust component will be much lower if an optically thin dust spectral index of 3 is used. 
This confirms that the continuum at 7 mm (and at lower frequencies) is indeed dominated 
by the free-free emission, and that the method of using a free-free continuum model to fit the 
VLA band continuum is valid.

\subsection{ALMA Source 3}
\label{sec:source3}

As shown in \S\ref{sec:HRL}, source 3 has strong H30$\alpha$ emission. 
The H30$\alpha$ flux integrated over area and velocity is 1.7 $\Jy~\kms$. 
Under the local thermal equilibrium (LTE) condition, such an H30$\alpha$ flux gives 
an optically thin limit of 1.3~mm free-free continuum ranging from $13~\mJy$ with $T_e=6,000~\K$ 
to $51~\mJy$ for $T_e=20,000~\K$ (Eq. \ref{eq:HRL_thin}). 
Such a free-free flux is higher than the observed total continuum flux at 1.3~mm, except for the low end of $T_e$. 
If the continuum is not optically thin, the intensity ratio between the recombination line 
(after subtracting continuum) and the free-free continuum becomes even lower than that estimated in Eq. \ref{eq:HRL_thin}, 
which leads to even higher free-free continuum flux. 
Therefore, the LTE condition for the H30$\alpha$ of this source may be in question, 
i.e., the observed bright HRL emission in this source might be enhanced by maser emission. 
To date, millimeter HRL masers have been reported in a modest number of sources 
(MWC349A, \citealt[]{Martin89,Jimenez13}; MonR2-IRS2, \citealt[]{Jimenez13}; and G45.47+0.05, \citealt[]{Zhang19}). 
Here we consider this to be another case of a possible millimeter HRL maser.

\begin{figure}
\begin{center}
\includegraphics[width=\textwidth]{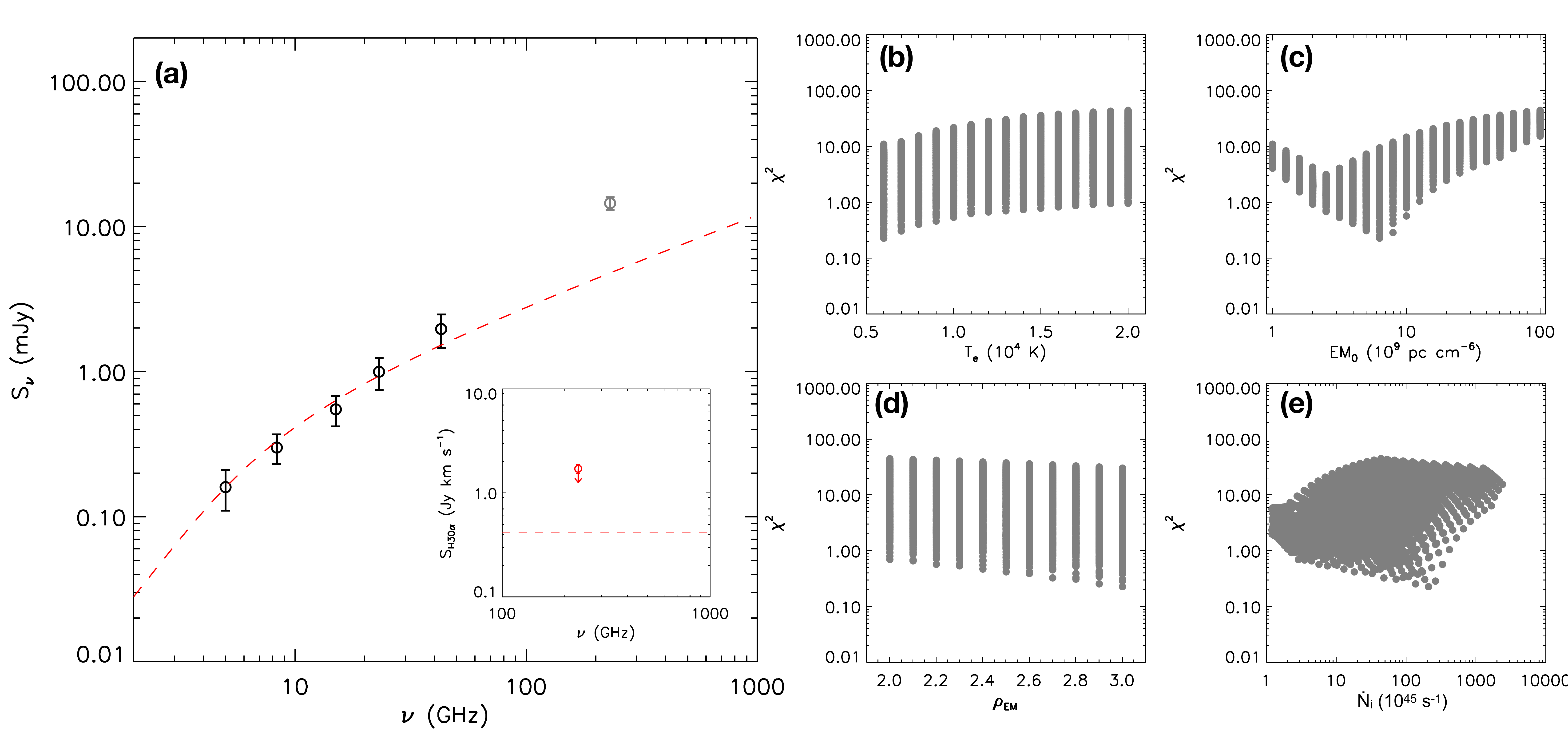}\\
\caption{Same as Figure \ref{fig:sed_source2}, but for ALMA source 3.
As the H30$\alpha$ line in this source has a potential maser component, we use it as an upper
limit to constrain the model (inset of panel a).}
\label{fig:sed_source3}
\end{center}
\end{figure}

As we do not have a non-LTE effect in our simple model, we use the observed H30$\alpha$ flux 
as an upper limit to constrain the model. 
Figure \ref{fig:sed_source3} shows the best-fit model and the $\chisq$ distribution in the parameter space. 
As neither the 1.3~mm continuum flux nor the H30$\alpha$ flux is used in fitting, 
we cannot constrain well the free-free emission at mm wavelengths. 
The emission measure is relatively well constrained to $\elecm=6.3\times 10^{9}~\pccm$ by the fitting, 
which leads to an electron density of $n_e\simeq1.1\times 10^7(l/{\mathrm{10au}})^{-1/2}~\mathrm{cm}^{-3}$
at the 10 au scale.
The compact size, the derived emission measure and electron density, 
as well as the H30$\alpha$ line width of 40~$\kms$ are also consistent with a HC{\Hii} region. 
The best-fit model has $T_e=6,000~\K$ and $\rho_\mathrm{EM}=3$, 
but the dependence of $\chisq$ on $T_e$ and $\rho_\mathrm{EM}$ are relatively weak.
As Figure \ref{fig:sed_source3}e shows, the models with similarly good fit give a range of 
$\dot{N}_\mathrm{i}=(0.3-30)\times 10^{46}~\mathrm{s}^{-1}$,
which corresponds to a ZAMS stellar mass of 10.7 to 17.3 $M_\odot$ 
(\citealt[]{Davies11}), i.e., spectral type of B1$-$B0 (\citealt[]{Mottram11}).
The best model has $\dot{N}_\mathrm{i}=2\times 10^{47}~\mathrm{s}^{-1}$,
corresponding to a ZAMS stellar mass of 16.4 $M_\odot$.

\subsection{ALMA Source 1}
\label{sec:source1}

\begin{figure}
\begin{center}
\includegraphics[width=\textwidth]{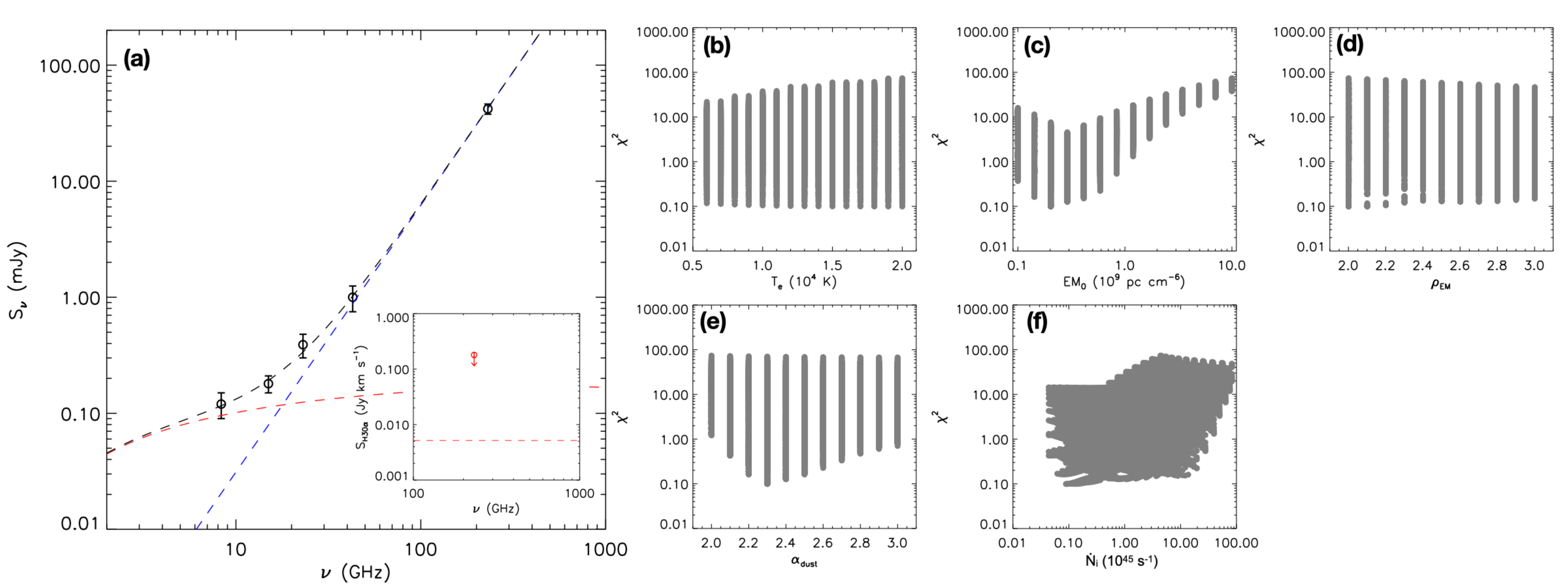}\\
\caption{{\bf (a):} The continuum spectral energy distribution of source 1
for VLA bands and ALMA 1.3~mm continuum (data points).
The red dashed line shows the free-free emission of the best-fit model.
The blue dashed line shows the dust emission of the best-fit model.
The black dashed line shows the total continuum emission.
The inset panel shows the $3\sigma$ level of the H30$\alpha$ integrated map 
used as upper limit (red data point) and the model prediction (red dashed line).
{\bf (b)$-$(e):} $\chisq$ distribution with model parameters: electron temperature
 $T_e$ (panel b), the emission measure $\elecm_0$ at the radius of $10~\au$
 (panel c), the power-law index of the emission measure distribution $\rho_\mathrm{EM}$ (panel d),
 and the spectral index of the dust component
 $\alpha_\mathrm{dust}$ (panel e).
 {\bf (f):} $\chisq$ distribution with the ionizing photon rate $\dot{N}_\mathrm{i}$
 calculated from $T_e$, $\elecm_0$, and $\rho_\mathrm{EM}$.}
\label{fig:sed_source1}
\end{center}
\end{figure}

No H30$\alpha$ emission is detected in the ALMA source 1 (core A or VLA source 2). 
The noise level of the integrated H30$\alpha$ emission map is 0.06 $\Jybeam~\kms$. 
We use the $3\sigma$ level (0.18 $\Jy~\kms$) as the upper limit for the H30$\alpha$ flux for this source. 
However, our free-free continuum and H30$\alpha$ model failed to reproduce the VLA band continuum fluxes 
and the H30$\alpha$ upper limit. 
The fact that the H30$\alpha$ emission is weak and the spectral index between 1.3~mm and 7 mm is greater than 2 
(see Figure \ref{fig:sed}), indicates that the dust emission may be significant even at 7 mm. 
Therefore, we include an additional dust component in the SED fitting (Eq. \ref{eq:dust}). 
We assume that the ALMA 1.3~mm continuum is dominated by the dust emission and use the 
dust spectral index $\alpha_\mathrm{dust}$ as an additional free parameter. 
The best-fit model and the $\chisq$ distribution in the parameter space are shown in Figure \ref{fig:sed_source1}. 
Note that since we only have five data points and one upper limit to constrain four free parameters, the constraints are relatively weak.

The emission measure and the dust spectral index are relatively well constrained. 
The best-fit model has $\elecm_{\mathrm{10au}}=2\times 10^{8}~\pccm$ 
(corresponding to an electron density of $n_e\simeq2.0\times 10^6(l/{\mathrm{10au}})^{-1/2}~\mathrm{cm}^{-3}$) 
and $\alpha_\mathrm{dust}=2.3$. 
On the other hand, the electron temperature $T_e$ and the emission measure gradient power-law index 
$\rho_\mathrm{EM}$ are not constrained. 
The models with similarly good fit give an ionizing photon rate range of $\dot{N}_\mathrm{i}=(0.09-10)\times 10^{45}~\mathrm{s}^{-1}$, 
which corresponds to a ZAMS stellar mass of 7.3 to 12.0~$M_\odot$ and a spectral type of B2 to B0.5. 
Interestingly, the derived emission measure and electron density are not very consistent with typical HC{\Hii} regions.
However, the compact size seen at radio wavelengths, the low free-free continuum flux, 
and the derived low central stellar mass, suggest that this is an extremely young photoionization region.

\subsection{ALMA Source 4}
\label{sec:source4}

\begin{figure}
\begin{center}
\includegraphics[width=\textwidth]{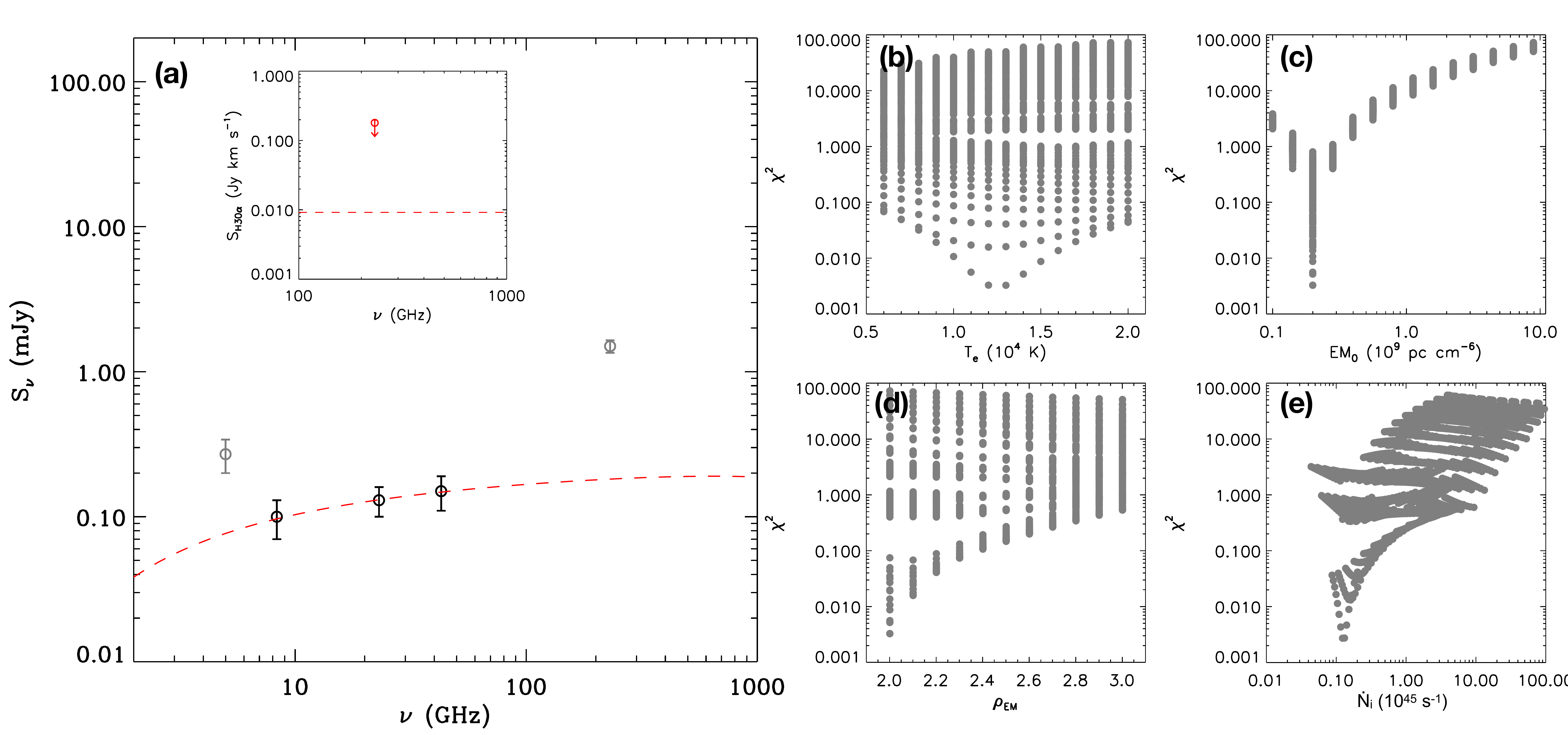}\\
\caption{Same as Figure \ref{fig:sed_source2}, but for ALMA source 4.
The 6~cm flux is also not used in the fitting (see the text).}
\label{fig:sed_source4}
\end{center}
\end{figure}

For source 4, the SED in the VLA frequency range is almost flat, 
indicating nearly optically thin free-free emission. 
Adopting the same method used for sources 2 and 3, we fit the three VLA bands 
with constraints for the upper limit of the H30$\alpha$ integrated flux ($3\sigma=0.18~\Jy~\kms$). 
The results are shown in Figure \ref{fig:sed_source4}. 
Note that we do not include the 6 cm continuum in the fitting, 
as it is significantly higher than the other VLA continuum fluxes (see Figure \ref{fig:sed}d),
which may be caused by non-thermal emission, which has negative indices. 
The ALMA 1.3~mm continuum is also not used because it has a significant contribution from dust emission. 
The best-fit model has $T_e=12,000~\K$, $\elecm_{\mathrm{10au}}=2\times 10^{8}~\pccm$, 
and $\rho_\mathrm{EM}=2$. 
These parameters lead to an ionizing photon rate of $\dot{N}_\mathrm{i}=1.3\times 10^{44}~\mathrm{s}^{-1}$, 
which corresponds to a ZAMS stellar mass of 7.6~$M_\odot$ and spectral type of B2.

However, as discussed in \S\ref{sec:continuum_highres},
while source 4 could be a forming massive star creating a 
small ionized region via photoionization like the fitting model assumes,
its low 1.3~mm continuum emission (on the same level as other low-mass sources
in the region) suggests that it could also be a low-mass protostar
with free-free emissions caused by shock-induced ionization from an unresolved jet.
In fact, the negative spectral index in the low-frequency range could indicate contribution of non-thermal emission
caused by the jet shock.
{\color{black} Another supporting evidence is that if the free-free emission is caused by photoionization,
the mass ratio between sources 2 and 4 is 2 to 3 as estimated above.
 If sources 2 and 4 are indeed a binary system, the existence of such a massive companion
should have significant impact on the circumbinary disk structure.
However, as we will show in \S\ref{sec:disk}, the disk structure is highly symmetric with respect to
source 2 indicating little to no impact from source 4}.

\subsection{Consistency of Stellar Mass Estimation}
\label{sec:mass}

In the above discussion, we estimate the stellar masses from the ionizing photon rates $\dot{N}_\mathrm{i}$
using the ZAMS stellar model, in which a value of $\dot{N}_\mathrm{i}$ corresponds to a single value
of the stellar mass $m_*$.
Here, we discuss the uncertainty of this method, comparing it to the dynamical masses and the
constraints from infrared luminosity.

{\color{black} A dynamical mass of 18 $M_\odot$ was derived by fitting the molecular line kinematics 
of the rotational structure around sources 2 and 4 (\citealt[]{Sanchez13}).}
Using the ZAMS ionizing rate, we estimate a total mass of about $18-26~M_\odot$,
which is close to, but slightly larger than, the dynamical mass of 18 $M_\odot$.
This difference can be completely due to uncertainties in both methods of mass estimation.
The mass estimated from the ionizing rate can be lower if some fraction of the free-free 
emission in source 4 is not caused by photoionization, as discussed above.
In that case, the total mass would be more consistent with the dynamical mass.
For source 1, previous studies of molecular lines provided dynamical mass of about 4 $M_\odot$
{\color{black} assuming the emission is from a Keplerian disk (\citealt[]{Sanchez14}).}
As shown in Section \ref{sec:vel_source1}, we improve the estimation of dynamical mass
to about $6~M_\odot$ as the source 1 disk is not rotationally supported.
This new dynamical mass is close to the mass evaluated from the ZAMS ionizing rate ($7.3-12~M_\odot$).

Next, we compare the infrared luminosity and the expected luminosity from the ionizing photon rate and the ZAMS model.
As the sources are embedded in the opaque dusty cloud, infrared observations provide only the total luminosity rather 
than the individual ones.
Under the assumption of the isotropic radiation, 
the total bolometric luminosity of the region is 
$L_\mathrm{bol,iso}\approx 3\times 10^4~L_\odot$, found by integrating the
infrared SED (e.g., \citealt[]{Sanchez14}).
More accurately, the SED fittings, which take into account the photon escape along the outflow cavity, 
gave the bolometric luminosity of $L_\mathrm{bol}\simeq(4-22)\times 10^4~L_\odot$ (\citealt[]{Zhang13,Debuizer17,ZT18}).
Although a single protostar was assumed in these fittings,
the obtained total luminosity is reasonable for the multiple-source system
as the ratio between $L_\mathrm{bol}$ and $L_\mathrm{bol,iso}$ mostly depends on the distribution of the surrounding material.
On the other hand, Table \ref{tab:sedmodel} 
lists the expected luminosities from the ionizing photon rates and the ZAMS model for each source.
Their total of $(1.8-6.8)\times 10^4~L_\odot$ is within the consistent range of the bolometric luminosity $L_\mathrm{bol}$ from the model fittings,
suggesting that our mass estimation from the ionizing rates is reasonable.
We note again that, however, an accreting massive protostar tends to have a bloated cool 
photosphere emitting fewer ionizing photons than the ZAMS star with the same luminosity
\citep{Hosokawa09,Hosokawa10,Haemmerle16,Tanaka16}.
It could explain why the total expected luminosity from the ZAMS ionizing rate (Table 2) is possibly lower than $L_\mathrm{bol}$.

In summary, the mass estimates from our free-free and H30$\alpha$ model fits are 
generally consistent with other constraints from kinematics and infrared luminosity.
A more accurate mass estimation should simultaneously take into account all the constraints discussed above, 
including ionizing photon rate, bolometric luminosity and kinematics, using realistic massive protostellar evolution models
rather than simple ZAMS stellar models.

\section{Molecular Line Observations}
\label{sec:line} 

Our observations include various molecular lines tracing different gas structures in the region. 
Here, we present results for the SO$_2$($22_{2,20}-22_{1,21}$), H$_2$CO($3_{2,1}-2_{2,0)}$), 
CH$_3$OH($4_{2,2}-3_{1,2}; E$), H$_2$S($2_{2,0}-2_{1,1}$), 
and CH$_3$OCH$_3$($13_{0,13}-12_{1,12}$) lines which
trace dense gas, and $^{12}$CO(2-1) and SiO($5-4$) which trace outflows.
The observing parameters for these lines are listed in Table \ref{tab:lines}.
{\color{black} These lines are the main targeted lines of our spectral window setups.
While some other weaker lines are also detected in the same spectral windows, we defer the 
analysis of the full molecular line data to a future study.} 

\subsection{Multi-scale Accretion Structures}
\label{sec:disk}

\begin{figure}
\begin{center}
\includegraphics[width=\textwidth]{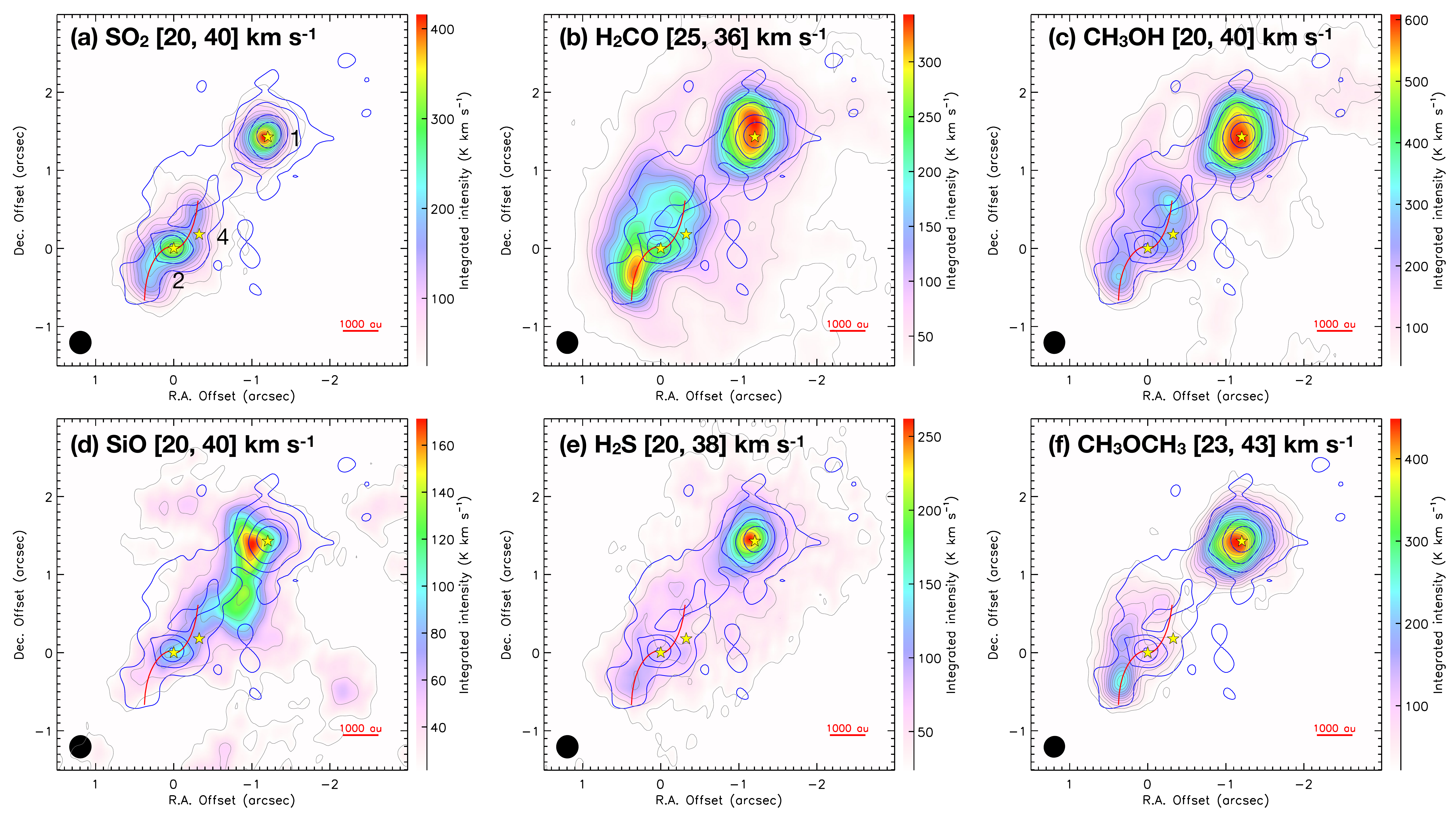}\\
\caption{Integrated intensity maps of different emission lines (color scale and grey contours)
overlaid with the 1.3~mm continuum emission (blue contours).
The molecule names and integrated $\vlsr$ ranges are labeled in each panel.
The continuum contour levels are at $10\sigma\times2^n$ 
($1\sigma=0.48~\mJybeam$, $n=0,1,...$). 
The line emission contours have lowest level and intervals of $3\sigma$,
with $1\sigma=8.3$, 7.6, 12.5, 7.3, 8.0, and 7.2 $\K~\kms$ in panels a$-$f, respectively.
The yellow stars mark the positions of sources 1, 2, 4 identified from the C9 configuration data
(source names are labeled in panel a).
The red curve shows the spiral structure fitted from the SO$_2$ integrated map.
}
\label{fig:intmap}
\end{center}
\end{figure}

Figure \ref{fig:intmap} shows the integrated emission maps of the 
SO$_2$,  H$_2$CO, CH$_3$OH, SiO, H$_2$S, and CH$_3$OCH$_3$ lines. 
Here we focus on the molecular lines associated with the central sources, 
especially source 1 (core A), and sources 2 and 4 (core B). 
For source 1, the line emission is mostly concentrated near the source center 
(except for the SiO emission whose peak is more offset). 
However, these lines show different spatial distributions around source 2.
{\color{black} Note that sources 2 and 4 are proposed to form a binary system by
\citet[]{Beltran16} based on their closeness and mass estimates.
However, it is still unclear whether they are a true bound binary system, or how they
relate to the nearby other sources identified in 1.3~mm long-baseline data.}

The SO$_2$ emission around source 2 is concentrated toward the center, 
and shows a morphology consistent with that of two spiral structures that are 
symmetric with respect to the center (Figure \ref{fig:intmap}a). 
In contrast to the SO$_2$ line, the center is devoid of H$_2$CO and CH$_3$OH emissions. 
While the spiral structures are still distinguishable in the
 H$_2$CO and CH$_3$OH emission, 
they mainly show an 
elliptical ring around sources 2 and 4.
In addition, the H$_2$CO and CH$_3$OH lines further show a stream of emission to the north of source 2 that connects to source 1,
forming a larger elliptical shape surrounding sources 1 and 2.
The aspect ratio of this larger elliptical structure is about 2:1. 
If we assume that this structure is circular, then the inclination between its midplane and the line of sight is about 30$^\circ$. 
This is close to the derived inclination of the disk around source 2 estimated from previous observations (\citealt[]{Sanchez13}). 
Therefore, despite the fact that material is distributed in the filament-like structure on the larger scale, 
there may be a near-circular structure with a radius of about 4000 au around the two main forming 
massive stars (sources 1 and 2), which then feeds into two separate disks feeding each source. 
In addition, the disk feeding source 2 (and source 4) has formed spiral structures.

\begin{figure}
\begin{center}
\includegraphics[width=\textwidth]{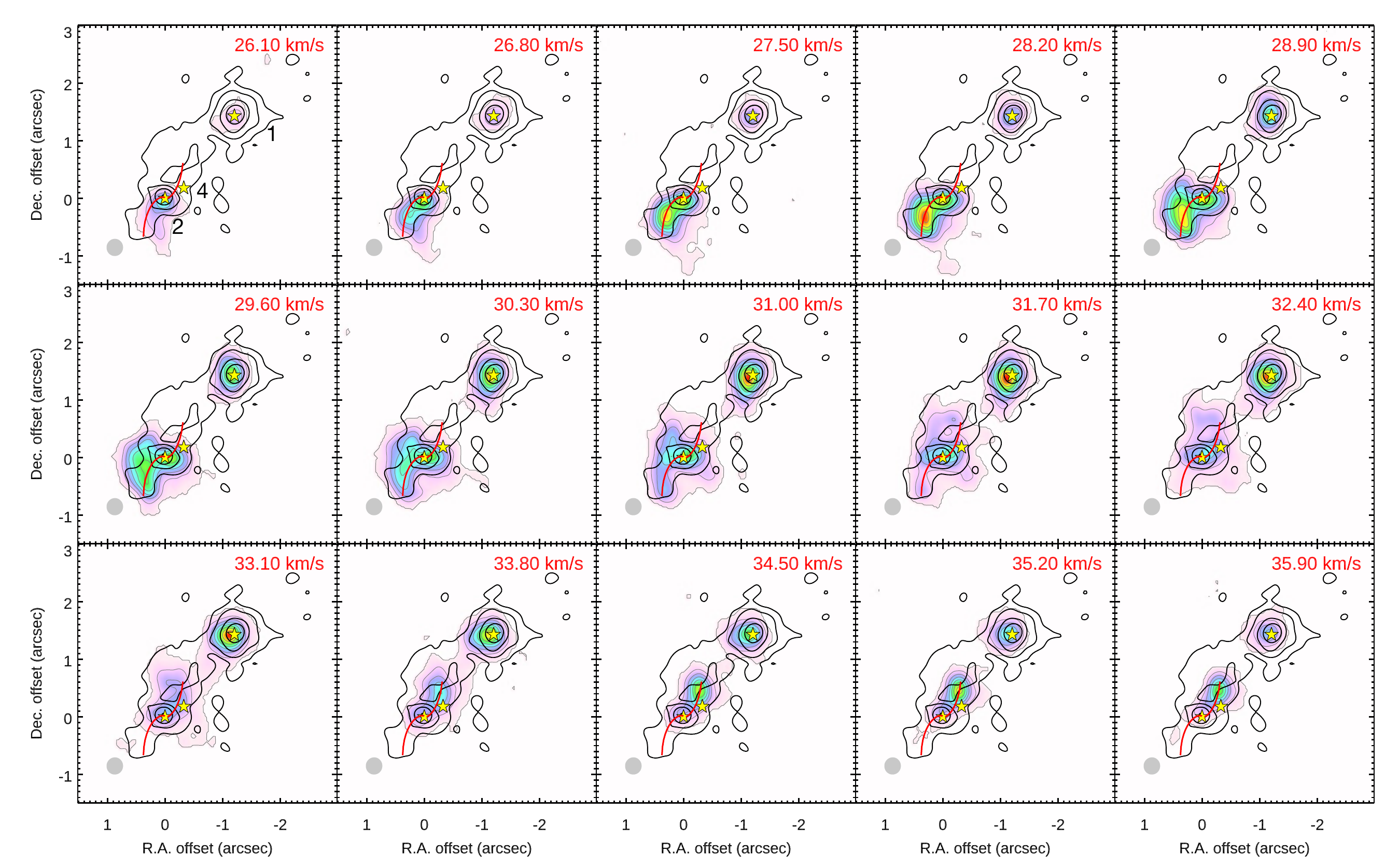}\\
\caption{Channel maps of the SO$_2$ ($22_{2,20}-22_{1,21}$) emission (color scale and grey contours).
The central velocity of each channel is labeled in each panel. 
The grey contours start at $3\sigma$ and have intervals of $6\sigma$ ($1\sigma=3.3~\mJybeam$).
The black contours show the continuum emission (same as Figure \ref{fig:intmap}).
The yellow stars mark the positions of sources 1, 2, 4 identified from the C9 configuration data
(source names are labeled in top-left panel).
The red curve show the spiral structure fitted from the SO$_2$ integrated map.
The synthesized beam is shown in the lower-left corner of each panel.
}
\label{fig:chan_SO2}
\end{center}
\end{figure}

\begin{figure}
\begin{center}
\includegraphics[width=\textwidth]{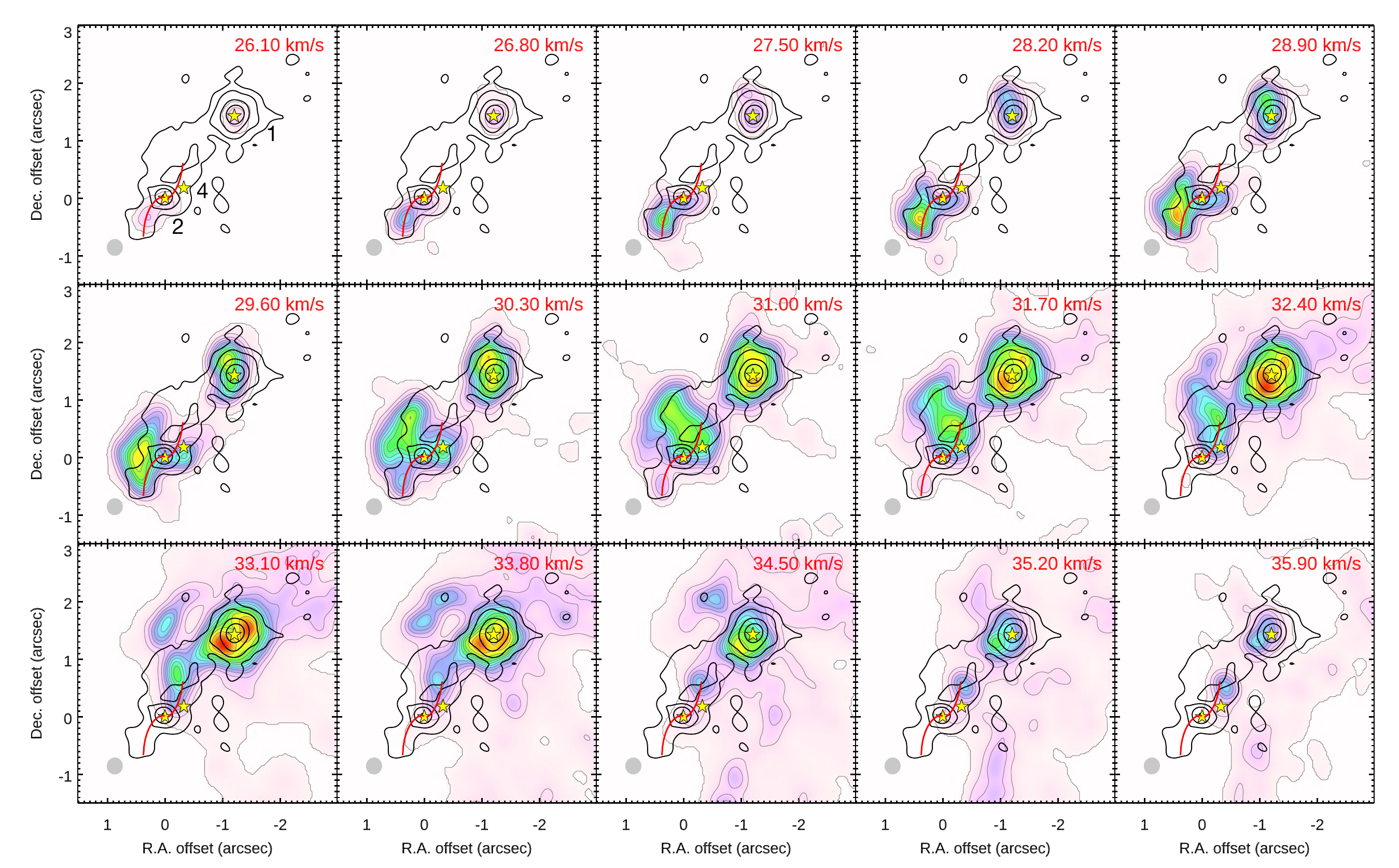}\\
\caption{Same as Figure \ref{fig:chan_SO2}, but for the CH$_3$OH 
($4_{2,2}-3_{1,2}; E$) line. 
The grey contours start at $3\sigma$ and have intervals of $6\sigma$ ($1\sigma=3.7~\mJybeam$).}
\label{fig:chan_CH3OH}
\end{center}
\end{figure}

The kinematics of these lines are consistent with such a scenario. 
Figures \ref{fig:chan_SO2} and \ref{fig:chan_CH3OH} show the channel maps of the 
SO$_2$ and CH$_3$OH lines around these sources. 
The spiral structures around source 2 seen in the integrated maps have highly symmetric and ordered velocity structures, 
with the southeastern arm blue-shifted and the northwestern arm red-shifted. 
Source 1 (core A) is slightly more red-shifted than source 2 (core B), and the stream connecting these two sources 
(Figure \ref{fig:chan_CH3OH}) is red-shifted with respect to source 2 and joins the source 1 disk at a velocity slightly 
more red-shifted than the source 1 systemic velocity.

Compared to the H$_2$CO and CH$_3$OH lines, the H$_2$S and CH$_3$OCH$_3$ emissions are more 
concentrated around the two main sources, but their morphologies are similar to the 
H$_2$CO and CH$_3$OH lines. 
SiO emission around source 2 appears to follow the spiral arm structures seen in the SO$_2$ line. 
However, SiO emission is widely spread out in this region
with multiple components tracing outflows (originated from source 1) and emission
along the filament, as we will discuss in more details in \S\ref{sec:outflow}.
Therefore, it is difficult to conclude that the SiO emission around source 2
actually traces the disk spiral structure.

\subsubsection{Velocity Structure around Source 2}
\label{sec:vel_source2}

Figure \ref{fig:pvdiagram} shows the position-velocity (PV) diagrams along a position angle 
of $\pa=-25^\circ$ across source 2 to show the kinematics of its {\color{black} proposed} disk in these molecular lines. 
This position angle is consistent with the position angle of the disk derived from 
the previous 0.85~mm observation at a lower angular resolution (\citealt[]{Sanchez13}). 
The spiral structures are also roughly aligned along this direction. 
Velocity gradients consistent with rotation are seen in all molecular lines. 
However, only in the SO$_2$ line, a high-velocity component reaching a 
rotation velocity of about 10 $\kms$ is seen within about 0.2$\arcsec$ (440 au) from the center. 
This high-velocity component is missing in all other molecular lines shown here. 
On the other hand, the low-velocity component of SO$_2$ is highly consistent with the other lines. 
The high-velocity inner component seen in the SO$_2$ line may be caused by an inner disk rotating faster.

There are two reasons why the SO$_2$ line can better trace the inner region and highlight the spiral structure. 
First, the observed SO$_2$ transition has a higher upper energy level {\color{black} ($E_u/k=248~\K$)} 
than the other lines shown here (see Table \ref{tab:lines}). 
Therefore it is more sensitive to the warmer inner region while the other molecular lines are more dominated by the 
outer cooler area. 
Second, as a typical shock tracer, the SO$_2$ emission may be enhanced by shocks associated with the spiral arms, 
and therefore more concentrated along the spiral arms than the other species. 
Such behavior is very similar to what has been reported in several other massive protostellar disks, 
such as G339.88$-$1.26 (\citealt[]{Zhang19a}) and IRAS16547$-$4247 (\citealt[]{Tanaka20}). 
In those cases, SO$_2$ lines are also found to trace inner region in the disk/envelope system with faster rotation than 
molecular lines such as CH$_3$OH and H$_2$CO.
However, in those sources, SiO emission is found to trace the innermost part of the disk with highest rotation velocities, 
which is not seen in the case of G35.2. 

\subsubsection{Fitting the Spiral Structures around Source 2}
\label{sec:spiral}

To better understand the properties of the spiral structure seen in the SO$_2$ line, 
we fit the them with two spiral shapes; one is a logarithmic spiral with the form of $R=R_0 e^{b\theta}$, 
and another is an Archimedean spiral with the form of $R=R_0+b\theta$ 
{\color{black} (e.g. \citealt[]{Huang18,Kurtovic18}).}

\begin{figure}
\begin{center}
\includegraphics[width=\textwidth]{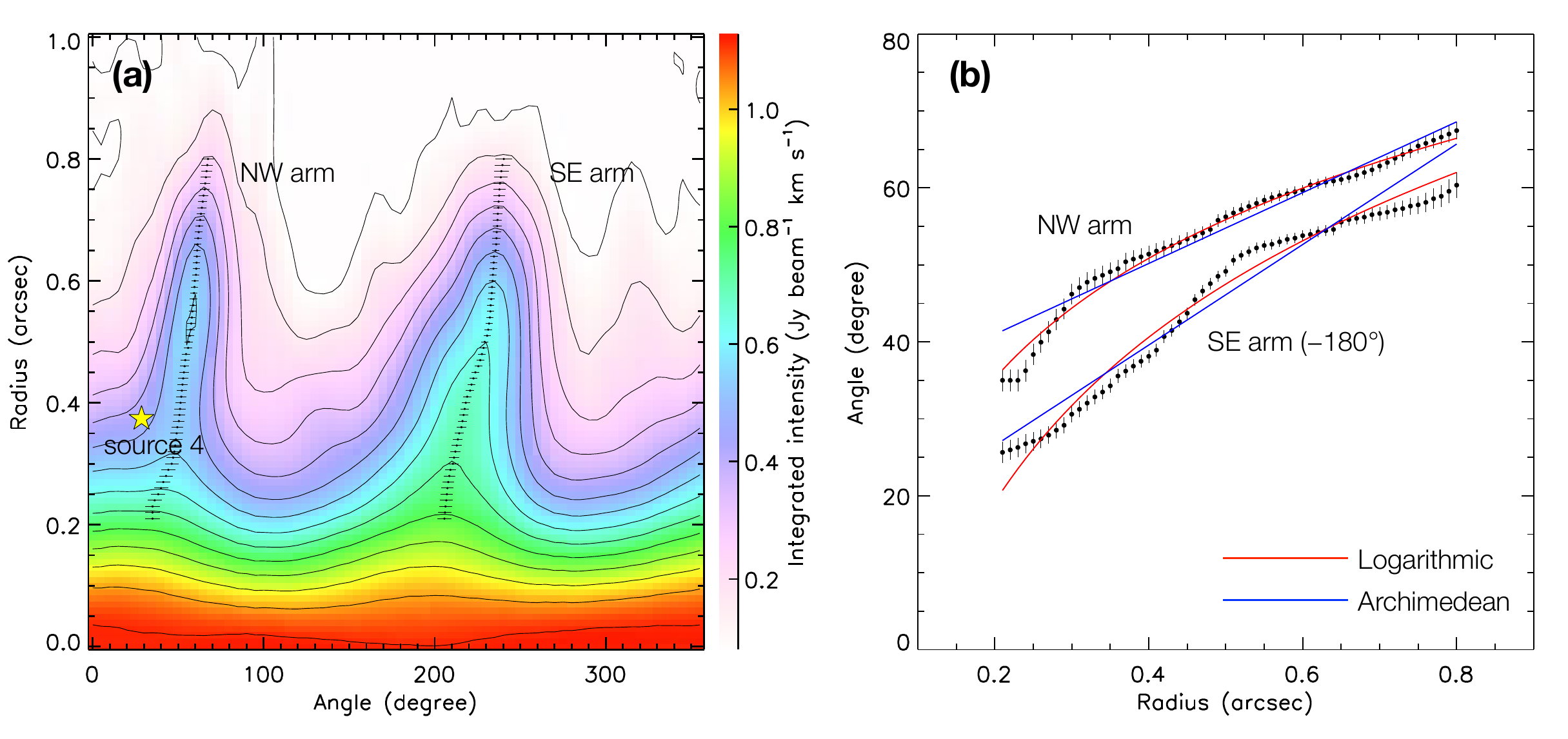}\\
\caption{{\bf (a)}: Distribution of the SO$_2$ integrated emission
in the space of radius from ALMA source 2 and polar angle.
The yellow star marks the location of source 4.
For each radius between 0.2$\arcsec$ and 0.8$\arcsec$ where
the spiral structures can be well defined, the polar angles
of the emission peaks of the two spirals are determined (data points).
The errors {\color{black} (shown in $3\sigma$ levels)} are determined by the beam size, radius and the S/N,
following $\Delta \theta=\theta_\mathrm{beam}/(2r~\mathrm{S/N})$,
where $\theta_\mathrm{beam}=0.3\arcsec$ and $r$ is the radius (distance
to source 2).
{\bf (b)}: Fits to the spiral structures in the SO$_2$ integrated emission map
around ALMA source 2.
The polar angle of the SE arm are rotated by $180^\circ$ to compare with the NW arm.
Each spiral structure is fitted with a logarithmic spiral (red curve) and
an Archimedean spiral (blue curve).
}
\label{fig:spiral}
\end{center}
\end{figure}

Figure \ref{fig:spiral}(a) shows the integrated SO$_2$ emission in terms of distance from source 2 and polar angle. 
We determine the polar angle of the two spiral arms by their emission peaks at each distance on the emission polar map. 
Only the radius range between 0.2$\arcsec$ and 0.8$\arcsec$ is used as the spiral structure is most clear in this range. 
Figure \ref{fig:spiral}(b) shows the fit to the determined spiral shapes. 
The parameters of the best-fit shapes are listed in Table \ref{tab:spiral}. 
The pitch angles of the spirals $\phi=\tan^{-1}(\frac{dR}{Rd\theta})$ are also listed. 
{\color{black} It appears that the logarithmic shape fits the spiral arms better than the Archimedean shape,
reproducing the curved shape (also see Table \ref{tab:spiral}).}
The pitch angle is $69^\circ$ for the NW arm and $62^\circ$ for the SE arm.
Note that the two spiral arms lie along the major axis of the disk,
the large pitch angles are partly due to the high inclination (closer to edge-on view) and the low spatial resolution.

There are commonly two mechanisms to form the spiral structures in a disk, 
gravitational instability and perturbation by a companion or a planet (e.g., \citealt[]{Huang18,Forgan18}). 
A pair of symmetric logarithmic spiral arms are expected if the spiral structures are formed due to gravitational 
instability in the disk (e.g., \citealt[]{Forgan18}). 
Also, the disk around source 2 is estimated to be about $3~M_\odot$ (\citealt[]{Sanchez13}), 
which is relatively massive with respect to the mass of the central sources 
(sources 2 and 4 have a dynamical mass of about $18~M_\odot$ in total; see \S\ref{sec:mass}). 
Such a massive disk is prone to gravitational instability. 
Another possibility is that the spiral structures are caused by the perturbation of the binary companion, i.e. source 4.
There are also several lines of evidence supporting such a scenario. 
First, the NW spiral arm shows a local minimum close to the location source 4 (see Figure \ref{fig:spiral}a),
which may indicate that the binary companion has opened a gap in the spiral. 
Furthermore, despite that a single Archimedean spiral cannot well fit the NW arm, 
this can be fit with two Archimedean shapes with a discontinuity at $r\approx 0.3\arcsec$, 
which leads to a local maximum of pitch angle. In fact, simulations have shown that the spirals formed 
by perturbation of a companion have increasing pitch angle toward the position of the companion 
(e.g. \citealt[]{Zhu15,Bae18}), which is similar to what we observed in the NW arm. 
Therefore, with the current data, it is difficult to determine unambiguously 
which is the formation mechanism of the spiral structures in the source 2 disk.

\begin{figure}
\begin{center}
\includegraphics[width=\textwidth]{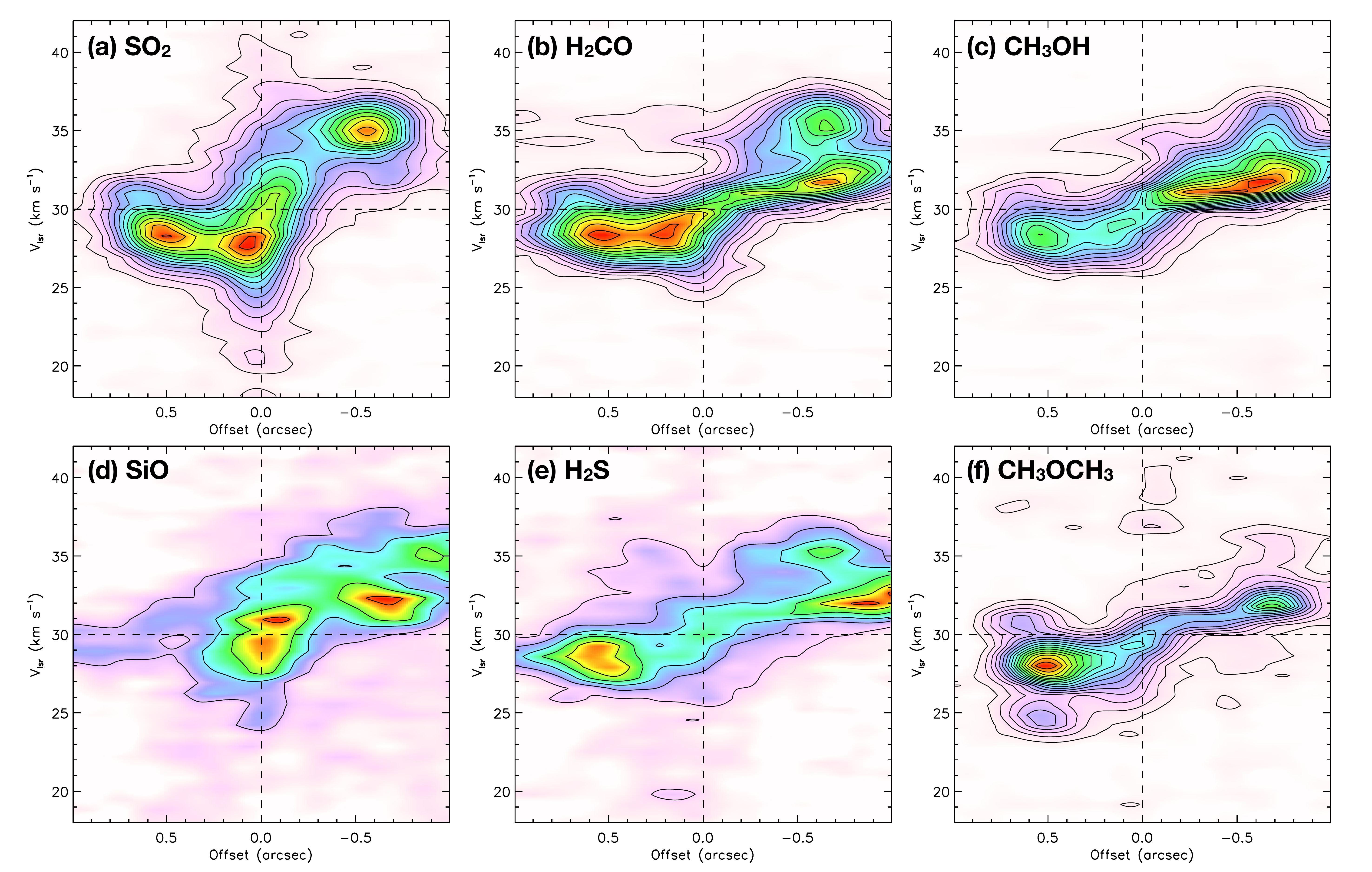}\\
\caption{Position-Velocity (PV) diagrams of the 
(a) SO$_2$, (b) H$_2$CO, (c) CH$_3$OH, (d) SiO, (e) H$_2$S, and (f) CH$_3$OCH$_3$
emission around source 2 along a cut with a position angle of $\pa=-25^\circ$, i.e. the disk orientation
obtained from the fit done by \citet[]{Sanchez13}, shown in color scales and black contours.
The position offset is relative to the continuum peak position of source 2 identified from the
C9 configuration data.
The horizontal line indicates the systemic velocity of $\vsys=30~\kms$ of the source,
The contours have lowest level and intervals of $3\sigma$,
with $1\sigma=3.2$, 3.7, 4.5, 3.3, 3.2, and 2.6 $\mJybeam$ in panels a$-$f, respectively.
}
\label{fig:pvdiagram}
\end{center}
\end{figure}

{\color{black} However, it is also possible that the rotation structure around source 2 
on the 2,000 au scale is not truly a disk. 
The extent of the spiral structures ($\sim 2,000~\au$) is much larger than the size of typical Keplerian disks 
around massive protostars seen in high-resolution observations,
which have typical radii $\lesssim$ a few $\times 10^2~\au$ 
(e.g. \citealt[]{Ginsburg18,Ilee18,Maud19,Zhang19b}).
Furthermore, although the kinematics of the rotation structure around source 2 was explained
by Keplerian rotation with lower-resolution data (\citealt[]{Sanchez13}),
its kinematics can also be explained by rotating-infalling material in the outer part feeding an inner small disk
(see Figure \ref{fig:pvdiagram}). The high-velocity component shown in the SO$_2$ PV diagram 
(Figure \ref{fig:pvdiagram}a) has a radius of $\sim 500~\au$, which is more consistent with other Keplerian disks
in massive star formation. 
If this is the case, the spiral structures can be the infalling streamers feeding the inner disk.
Such streamers on several $\times 10^3~\au$ scales were reported in other sources 
(e.g. \citealt[]{Pineda20}), and can be explained by accretion of unsmooth turbulent material or even cloudlets
(e.g. \citealt[]{Kuffmeier19,Hanawa22}). However, it is unclear whether such a scenario
can explain the high symmetry seen in the pair of the spiral structures.
}

\subsubsection{Velocity Structure around Source 1}
\label{sec:vel_source1}

\begin{figure}
\begin{center}
\includegraphics[width=\textwidth]{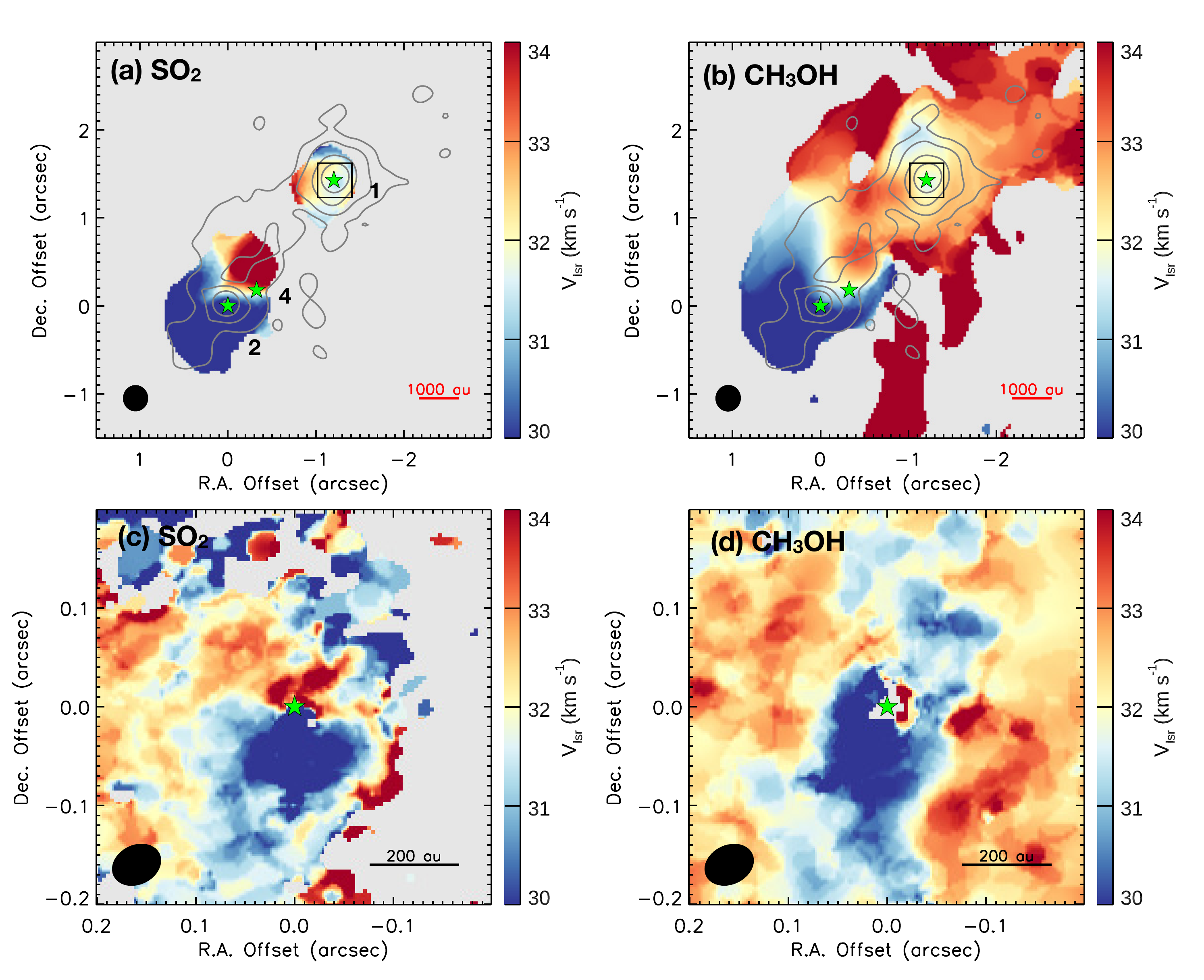}\\
\caption{{\bf (a)}: Moment 1 map of the SO$_2$ ($22_{2,20}-22_{1,21}$) emission shown
in color scale, overlaid by the continuum emission shown in grey contours.
The three stars mark the positions of sources 1, 4, 2.
The square indicates the region shown in panel b.
{\bf (b)}: same as panel a, but for the CH$_3$OH ($4_{2,2}-3_{1,2}; E$) line.
{\bf (c)}: A zoom-in view of the moment 1 map of the SO$_2$ emission around source 1 (marked with the star).
Data combining C3, C6 and C9 configurations are used.
{\bf (d)}: same as panel c, but for the CH$_3$OH ($4_{2,2}-3_{1,2}; E$) line.
}
\label{fig:velmap}
\end{center}
\end{figure}

Rotation velocity gradients are also seen around source 1.
{\color{black} In the low-resolution images}, the emission around source 1 appears to be blue-shifted 
in the {\color{black} northwest} and red-shifted in the {\color{black} southeast},
opposite from that around source 2 (see Figure \ref{fig:velmap}).
Similar velocity gradient directions were reported by \citet[]{Sanchez14} with ALMA 0.85 mm observations.
{\color{black} However, with higher resolution, the innermost region ($<0.1\arcsec$, $\sim200~\au$)
around source 1 shows a different velocity gradient direction, as shown by panels c and d in Figure \ref{fig:velmap}.
The innermost SO$_2$ emission is blue-shifted in the south and red-shifted in the north.
The CH$_3$OH line has blue-shifted emission in the southeast. There is some red-shifted emission
in the northwest of source 1 in the innermost region, but is affected by absorption.
The velocity gradients in the innermost region around source 1 revealed by these two molecules 
are opposite from that in the outer region 
but generally consistent with that seen around source 2.
This rotation direction is also consistent with the relative velocities of sources 1 and 2,
as source 1 (the northwestern source) is red-shifted with respect to source 2.}

\begin{figure}
\begin{center}
\includegraphics[width=\textwidth]{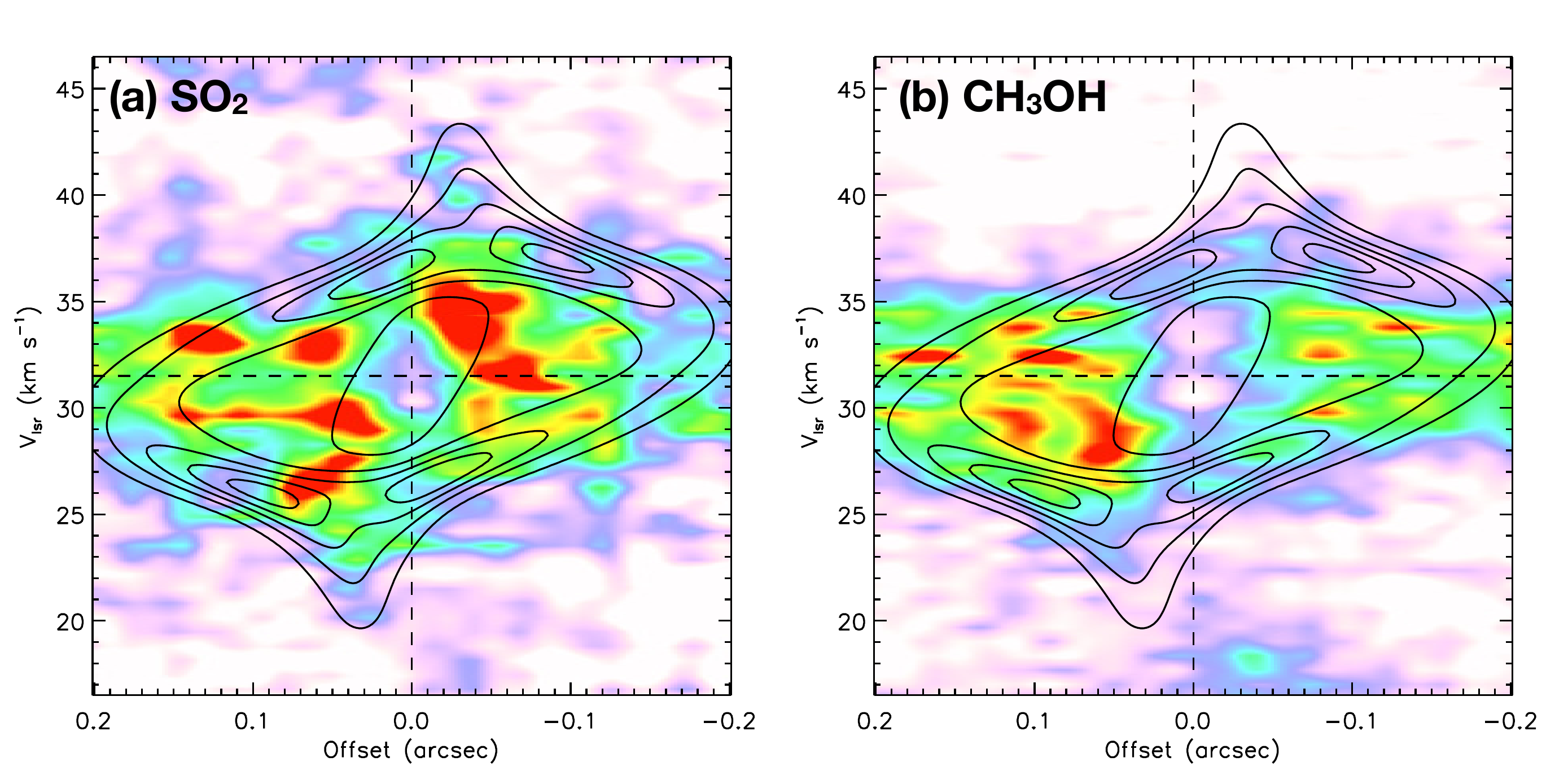}\\
\caption{{\bf (a)}: Position-velocity diagram of the SO$_2$ line along a direction with maximum velocity gradient across source 1
($\pa=0^\circ$, positive offset is toward south).  The combined data of C3, C6, and C9 configurations are used.
The black contours show the model of an infalling-rotating envelope (see the text).
{\bf (b)}: same as panels a, b, but for the CH$_3$OH ($4_{2,2}-3_{1,2}; E$) line.
The contours show the same model for the SO$_2$ emission.}
\label{fig:pvdiagram_source1}
\end{center}
\end{figure}

Figure \ref{fig:pvdiagram_source1} shows the position-velocity diagrams of SO$_2$ and CH$_3$OH lines,
along a direction with maximum velocity gradient across source 1 seen in the SO$_2$ moment 1 map 
($\pa=0^\circ$).
{\color{black} In the SO$_2$ PV diagram, 
there are higher-velocity components with clear gradients in the innermost region
that can be explained by rotation.}
However, in the case of pure rotation (e.g., a Keplerian disk),
the blue-shifted emission would be limited to one side while the red-shifted emission would be limited to the other side,
which is not compatible with the observed pattern.
Instead, a combined motion of infall and rotation can reproduce the observed kinematics.
{\color{black} Compared to the SO$_2$ emission, the CH$_3$OH emission shows similar
increase of velocity gradient toward the center. However, it does not show the central high-velocity component
as SO$_2$, indicating that the CH$_3$OH traces slightly outer regions
than SO$_2$, similar to the case of source 2 (\S\ref{sec:vel_source2}).
The CH$_3$OH line shows strong absorption against the bright continuum emission in red-shifted velocities
(see Figure \ref{fig:pvdiagram_source1}b). This indicates infalling material in front of the central source along
the line of sight, consistent with a combined motion of infall and rotation.}
Furthermore, if the accreting material is not evenly distributed (e.g., there are unsmooth accretion flows), 
the combined motion of infall and rotation can cause the change of velocity gradient directions from the
outer region to the inner region (e.g., \citealt[]{Liu17}).

{\color{black} To demonstrate the possibility of the combined motion of infall and rotation, we 
construct a simple model to fit the observed PV diagrams.}
Following the method used by \citet[]{Sakai14} and \citet[]{Zhang19b}, 
we fit the observed PV diagrams with a simple model of infalling-rotation
with the rotation velocity $v_\varphi$ and infall velocity $v_r$ described as
\begin{eqnarray}
v_\varphi (r) & = & \vcb \frac{\rcb}{r},\\
v_r (r) & = & -\vcb\frac{\sqrt{\rcb(r-\rcb)}}{r}.
\end{eqnarray}
Such motion conserves both angular momentum and mechanical energy.
$\rcb$ is the innermost radius that such infalling gas can reach with
the angular momentum conserved (i.e., the centrifugal barrier),
where $v_\varphi = \vcb$ and $v_r = 0$.
{\color{black} $\vcb$ is the rotational velocity at the centrifugal barrier.}
In such a model, the central mass is $m_{*}=\rcb\vcb^2/(2G)$.
Since we only focus on the kinematics pattern here, 
we adopt a simple geometry and density distribution in the
model. We assume the disk has a height of $h(r)=0.2r$ on each side
 of the mid-plane, and the density distribution follows $\rho(r)\propto
r^{-1.5}$ (e.g. \citealt[]{Oya16}).  For simplicity, we also assume the emissions are optically thin
and the excitation conditions are {\color{black} uniform} across the region.
{\color{black} Note that the density distribution, excitation conditions, and disk height profiles
have a combined effect on the detailed emission pattern, but do not affect the general kinematics pattern,
which we aim to explore here.
We also fix the outer radius to be $0.2\arcsec$ ($4.4\times 10^2~\au$),
as we only focus in the kinematics of the innermost region.
Therefore, we have in total three free parameters: 
the central mass $m_*$, the radius of the centrifugal
barrier $\rcb$, and the inclination angle $i$.}

To obtain the best-fit model, we compare the model PV diagram
with the observed PV diagram of the SO$_2$ line.
{\color{black} We focus on the SO$_2$ line as the CH$_3$OH line does not trace the innermost region well.}
We explore the inclination angle $i$
with values ranging from $0^\circ$ to $40^\circ$ with an interval of
$10^\circ$, the angular radius of the centrifugal barrier $\rcb/d$ in a range of
$0.01\arcsec - 0.05\arcsec$ with an interval of $0.01\arcsec$, and the
central dynamical mass in a range of $4-12~M_\odot$ with an interval of $1~M_\odot$,
and approximately determine the best-fit model by eye.
The SO$_2$ PV diagram can be well reproduced by the model
with $\rcb/d=0.03\arcsec$ ($\rcb\approx70~\au$), $i=30^\circ$ between the line of sight and the disk mid-plane,
$m_*=6~M_\odot$ (Figure \ref{fig:pvdiagram_source1}a).
{\color{black} In Figure \ref{fig:pvdiagram_source1}b, we show the comparison of this model and the 
CH$_3$OH emission. The CH$_3$OH emission is consistent with the model in the outer part, but
does not show the innermost high-velocity component. This confirms the scenario that CH$_3$OH
traces a region with slightly larger radii than that traced by the SO$_2$ emission.}


{\color{black} Our model fitting shows that infalling rotation can be a viable explanation for the gas kinematics 
in the innermost region around source 1. We note that we cannot rule out all the other possibilities.
Especially, our data have limited sensitivities for the extended emissions, and the rotation direction changes
from larger scale to the small scale, therefore our simple model cannot model the kinematics of more extended emission.}
Nevertheless, the model fitting suggests that the rotating structure around source 1 is not rotationally supported.
The derived central mass of $6~M_\odot$ is close to that estimated from the ionizing photon rate (see \S\ref{sec:spec}),
and higher than the previously derived dynamical mass of 4 $M_\odot$ (\citealt[]{Sanchez14}).
In the previous estimation, a Keplerian disk was assumed, higher dynamical mass
is expected if the disk is not rotationally supported.
Despite the non-detection of the rotationally supported disk in our observation, such a disk 
may exist inside the centrifugal barrier, i.e., within a radius of $0.03\arcsec$ ($\sim 70~\au$).
The resolution and line coverage of our current observation may not be able to effectively probe such a disk.

Overall, the hierarchical accretion structure around sources 1 and 2, 
in which both sources are fed by a common rotating structure,
and each of them is fed by the individual rotating structure, 
indicates that the accretion towards these two sources are still highly active and the start
of photoionization has not stopped the accretion.

\subsection{Molecular Outflows}
\label{sec:outflow}

\begin{figure}
\begin{center}
\includegraphics[width=\textwidth]{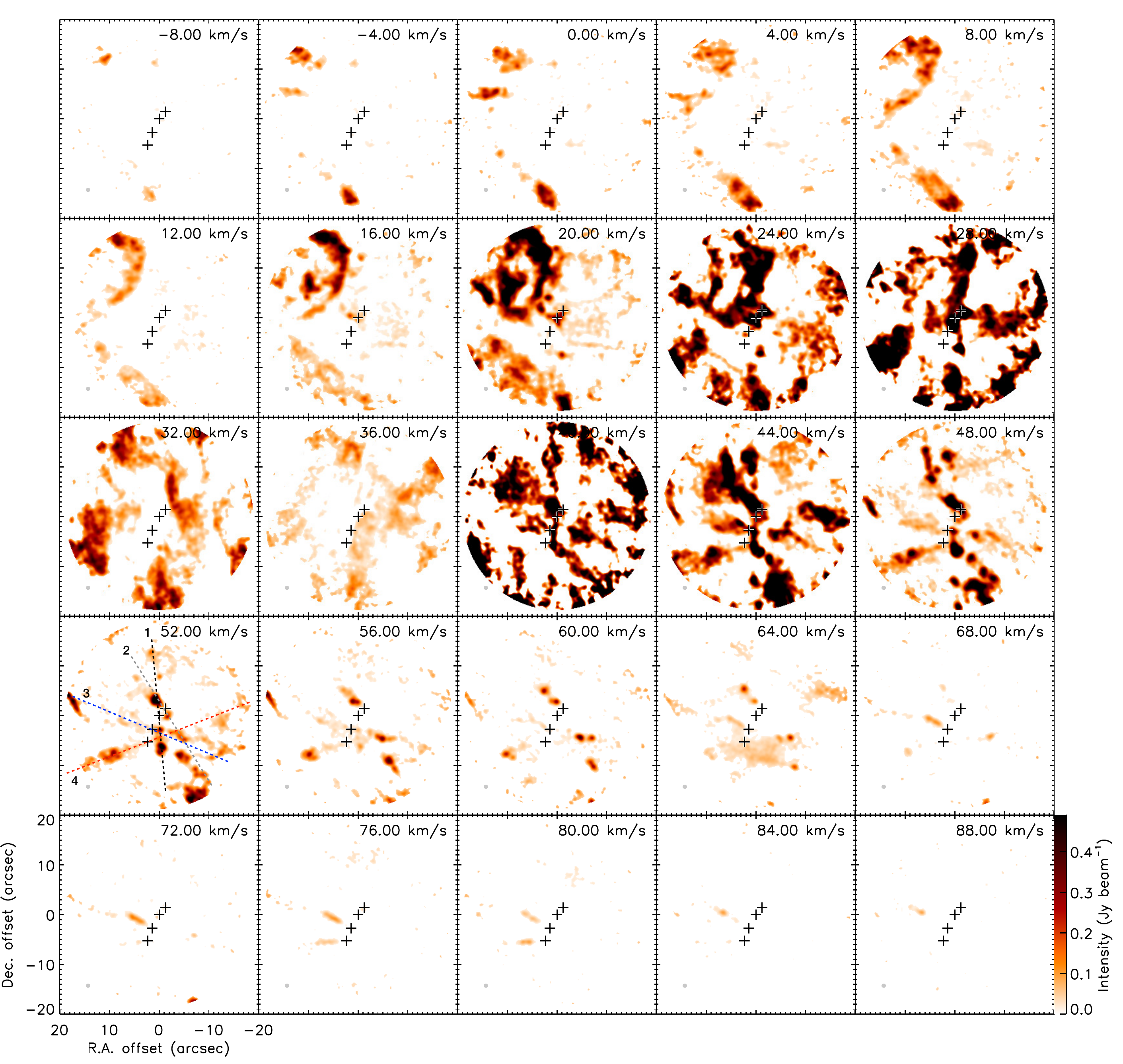}\\
\caption{Channel maps of the $^{12}$CO (2-1) line.
Each channel has a width of 4 $\kms$.
The central velocity of each channel is labeled in each panel. 
Only the C3 configuration data are used, 
and the synthesized beam is shown in the lower-left corner of each panel.
The dashed lines in the channel of $\vlsr=52~\kms$ show the directions
of the identified four outflow components.
The four crosses mark the four sources which we identify as the driving sources
of these outflow components.
From north to south, the four sources and their associated cores are source 1 (core A), 
source 2 (core B), {\color{black} source 14} (core C), {\color{black} source 17} (core E).}
\label{fig:chan_12CO}
\end{center}
\end{figure}

Figure \ref{fig:chan_12CO} shows the $^{12}$CO channel maps of G35.2. 
The $^{12}$CO emission shows a complex morphology. 
Our short observations with limited $u-v$ coverage cannot effectively recover the 
extended component of the emission, causing significant missing flux, 
which makes it even more difficult to analyze the outflows in this region.

\begin{figure}
\begin{center}
\includegraphics[width=\textwidth]{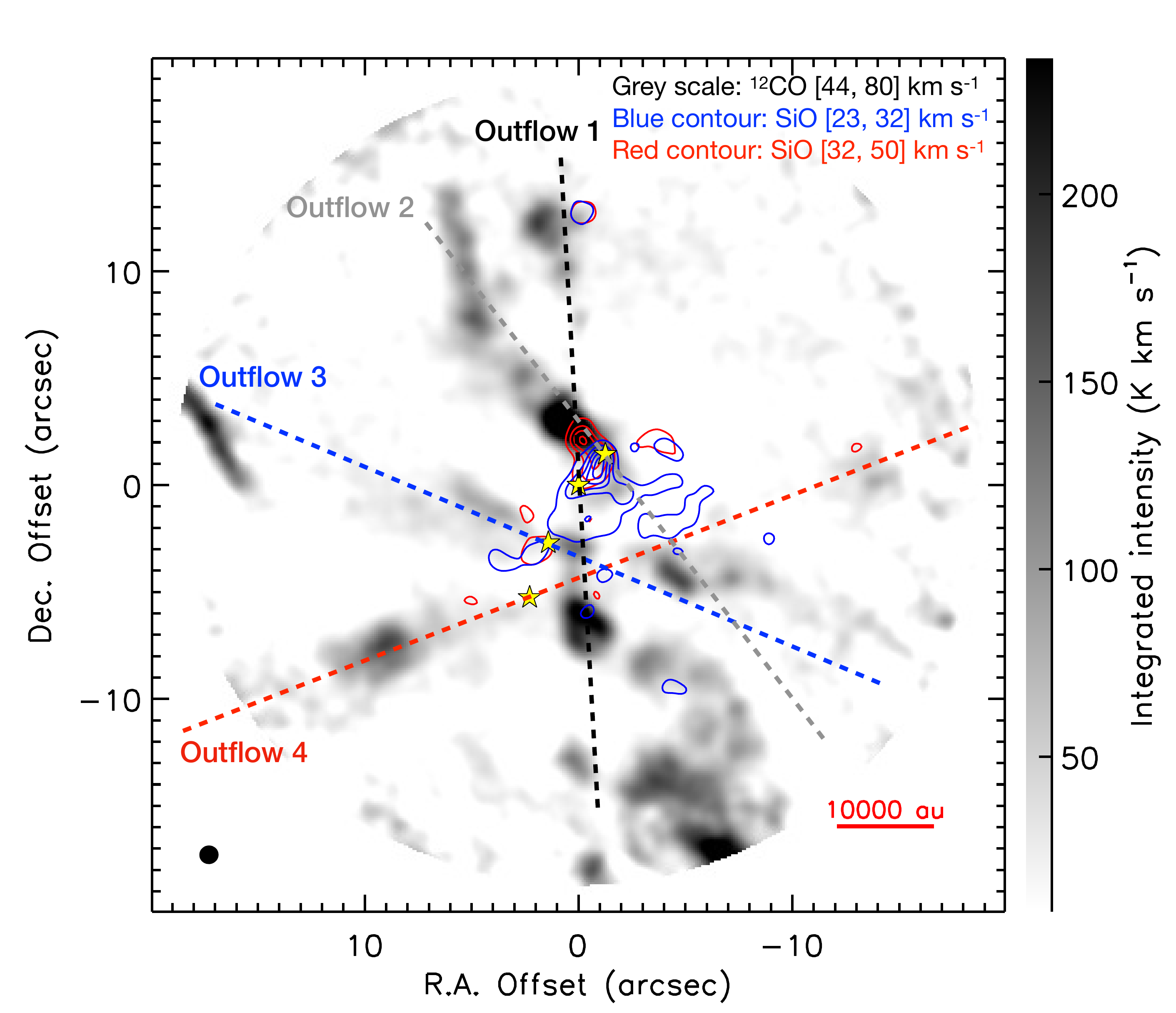}\\
\caption{Integrated map of the $^{12}$CO (2-1) line shown in grey scale,
overlaid with integrated maps of the blue-shifted and red-shifted SiO emission
shown in blue and red contours, respectively. The integrated velocity ranges
of the lines are shown in the upper-right corner of the figure.
The dashed lines show the directions
of the identified four outflow components (same as those in the channel of of $\vlsr=52~\kms$ in Figure \ref{fig:chan_12CO}).
The four crosses mark the four driving sources.
From north to south, the four sources and their associated cores are source 1 (core A), 
source 2 (core B), {\color{black} source 14} (core C), and {\color{black} source 17} (core E).}
\label{fig:intmap_12CO}
\end{center}
\end{figure}

In the blue-shifted velocities (with respect to the systemic velocity which is around 
$\vlsr=+30~\kms$ to $+35~\kms$; see \S\ref{sec:disk}), the most distinguishable outflow 
feature is a wide arc-shaped structure toward the northeast. 
This feature is best seen in velocity channels $\vlsr\lesssim+16~\kms$, 
but can also be seen in less blue-shifted velocities down to  $\vlsr=+24~\kms$. 
The wide-angle CO outflow towards NE was seen in previous observations, 
including single-dish observations (e.g., \citealt[]{Gibb03,Birks06}). 
This structure is also seen in the NIR both in molecular and atomic emission 
(\citealt[]{Caratti15,Fedriani19}).
On the other hand, in the red-shifted velocity channels, the emission appears to be more concentrated in a 
few collimated components, which are better shown in the integrated emission map shown in Figure \ref{fig:intmap_12CO}.
We identify in total four collimated outflow components, 
which are labeled by dashed lines with different colors in Figure \ref{fig:chan_12CO} (in the channel of $\vlsr=+52~\kms$)
and Figure \ref{fig:intmap_12CO}.
Assuming that they are separate outflows, we 
identify their driving sources by comparing their locations with the positions of the continuum sources.

Outflow 1 (marked with black dashed line) is most clearly seen in the channels of $+40~\kms\leq \vlsr \leq +52~\kms$, 
but it can also be seen in slightly blue-shifted channels of $\vlsr=+28~\kms$. 
At $\vlsr=+48~\kms$, it is clear that there is a string of bright CO blobs associated with this outflow. 
{\color{black} Despite the fact that this outflow component does not have a well defined structure in our data,
we believe it traces the same north-south outflow component identified in previous CO observations (e.g., \citealt[]{Birks06}), 
and consistent with the radio jet (\citealt[]{Gibb03,Beltran16}) and 
the bright outflow cavity seen in the NIR to MIR (\citealt[]{Fedriani19,Debuizer06,Zhang13}).}
This outflow appears to originate from source 2 (core B).

There appears to be an outflow associated with source 1 (core A), 
which is most clearly seen as two bright blobs to the northeast of source 1 at velocities $+56~\kms\leq\vlsr\leq +64~\kms$. 
This outflow has a position angle of about 30$^\circ$, if the blobs towards the opposite direction are also associated with it 
(marked with the grey dashed line in Figures \ref{fig:chan_12CO} and \ref{fig:intmap_12CO}).
This outflow is also traced by the SiO emission (Figure \ref{fig:intmap_12CO}), which is best seen
as a high-velocity red-shifted emission blob (up to $\vlsr>+50~\kms$) originating from source 1 to the same direction as
the $^{12}$CO outflow.
This SiO outflow has been reported by \citet[]{Sanchez14}. 
Its blue-shifted emission is more extended and mostly resides along the filamentary continuum structure.
It is possible that, in addition to the outflow, the SiO emission also trace the shocks caused by the gas flows
forming the filament.

Outflow 3 (marked with the blue dashed line) originates from {\color{black} source 14} (core C). 
It is clearly seen at velocities $\vlsr \geq +48~\kms$. 
To the northeast of {\color{black} source 14}, there is a high-velocity component in this outflow that can be 
seen from $\vlsr=+68~\kms$ up to $\vlsr>+90~\kms$. This outflow has a position angle of $70^\circ$. 
The position angles of Outflows 2 and 3 are highly consistent with the two larger-scale outflows 
seen in NIR shocked H$_2$ emission (\citealt[]{Caratti15}). 
Therefore, they are likely the same outflows seen in different wavelengths and different tracers. 
If this is the case, the wide CO outflow toward the northeast should be a combination of these two outflows 
rather than a single wide outflow. 
Furthermore, if these two outflows are driven by sources 1 and 15, 
they are separated from the North-South outflow (Outflow 1) driven by source 2. 
This suggests that the large NE outflow is not caused by the precession of the NS outflow/jet,
but rather, they are independent outflows driven by different sources.

Finally, we identify a fourth outflow (marked by a red dashed line in Figure \ref{fig:chan_12CO}) 
at a position angle of $110^\circ$ originated from {\color{black} source 17} (core E). 
This outflow is best seen to the east of {\color{black} source 17} at velocities of $+48~\kms\leq\vlsr\leq+60~\kms$, 
but can also be seen at high outflowing velocities ($\vlsr=+80~\kms$, with a slightly different position angle). 
This outflow has not been reported by previous studies.

{\color{black} Note that the outflows 3 and 4 are only seen clearly in the high-velocity red-shifted channels.
The crowded nature of this region, as well as as the fact that the observation is very shallow
with a lot of missing flux and side-lobe effects, 
make it very difficult to identify the different outflow structures in all the velocity channels.
Therefore, with the current data, it is hard to conclude if the corresponding blue-shifted components of outflows 3 and 4
are really missing or they are not seen because of the observation difficulties.}


\section{Lower-mass Members of the Cluster}
\label{sec:discussion}

As presented in \S\ref{sec:continuum}, we identify {\color{black} 22} compact sources from the ALMA long-baseline continuum image. 
As discussed in \S\ref{sec:spec}, the emission from 6 cm to 1.3 mm in sources 1$-$3 suggest that they 
are massive young stars that have created small photoionized regions around them.
Similar scenario may be also true for source 4, but it can also be a low-mass protostar with
shock-ionized region caused by jet activity. 
Other sources are assumed to be low-mass members in the forming cluster. 
Following previous similar studies (e.g., \citealt[]{Plunkett18,Busquet19}), 
we use several methods to characterize their distribution.

We assume that the detected {\color{black} long-baseline (C9)} 1.3~mm dust continuum emission 
in these sources comes from their disks {\color{black} and/or inner envelope (referred to as disks for simplicity)}, 
and therefore use the continuum fluxes to estimate {\color{black} the masses of such structures} of the 
sources using the following equation
\begin{equation}
M=\frac{d^2 F_\mathrm{dust,1.3mm}}{\kappa_\mathrm{1.3mm} B_\mathrm{1.3mm}(T_\mathrm{dust})},
\end{equation}
where $d=2.2$ kpc is the distance to the source, $F_\mathrm{dust,1.3mm}$ is the estimated dust continuum emission flux, 
and $B_\mathrm{1.3mm}(T)$ is the Planck function. 
An opacity of $\kappa_\mathrm{1.3mm}=0.00899~\mathrm{cm}^2~\mathrm{g}^{-1}$ is used 
(\citealt[]{Ossenkopf94}; for MRN dust with thin ice mantles and gas density of $10^6~\mathrm{cm}^{-3}$;
a gas-to-dust mass ratio of 100 is assumed). 
The disk masses are listed in Table \ref{tab:cont}. 
For sources 1$-$4 (massive sources), a dust temperature of 100 K is assumed
{\color{black} (suitable for typical massive sources; e.g. \citealt[]{ZTH14})}, 
and for sources 5$-$22, a dust temperature of 30 K is assumed {\color{black} (e.g. \citealt[]{Tobin13})}. 
Note that for sources 1$-$4, the free-free component estimated from modeling the SEDs and HRL fluxes 
(see \S\ref{sec:spec}) has been subtracted from the continuum fluxes. 
As the continuum fluxes are estimated with only the long-baseline image, 
the larger filament or accretion structure is filtered out, i.e., only the individual disk {\color{black} (and/or inner envelope)} 
of each source is included in the mass estimation.
{\color{black} Based on the assumed dust opacity, gas-to-dust ratio, and dust temperature of 30~K, 
the column density sensitivity is $6.3~\gcm$ at the center of the field of view, 
which corresponds to a mass sensitivity of $3.8\times 10^{-3}~M_\odot$
for unresolved sources.}

\begin{figure}
\begin{center}
\includegraphics[width=\textwidth]{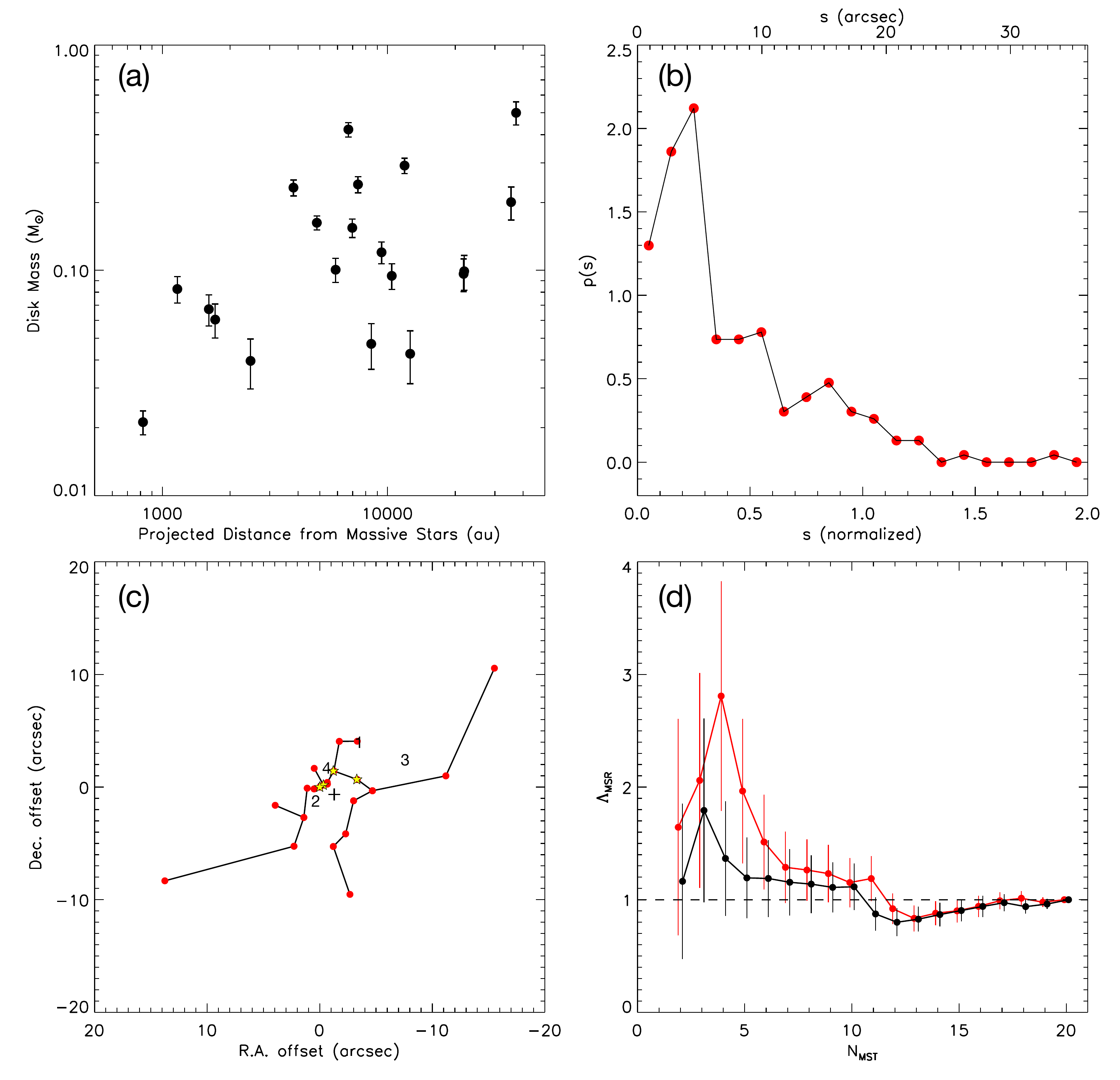}\\
\caption{{\bf (a)}: Disk mass (estimated from the total dust continuum flux)
of the compact sources detected with ALMA long-baseline observations 
(see Table \ref{tab:cont}) as a function of their projected distances from
the massive stars. Here, for any source, we calculate its distances to sources 1 and 2,
and use the smaller value as its distance from massive stars.
For sources 1$-$4, free-free emission is subtracted from the continuum 
(see Table \ref{tab:cont} and \S\ref{sec:spec}).
{\bf (b)}: Distribution function of separations between the ALMA sources.
 The quantity $p(s)\Delta s$ {\color{black} (area under the curve)} 
 is the number of pairs of sources with their separation falling
 within the range $[s,s+\Delta s]$.
 Separation $s$ is normalized to the cluster radius 
 {\color{black} (18.1\arcsec, $4.0\times 10^4~\au$)}, which is determined by
 the maximum separation between any source to the average position of all
 the sources.
 {\bf (c)}: Minimum spanning tree (MST) for the ALMA sources. 
The position offsets are with respect
to the position of source 2.
The stars mark the four massive sources with corresponding VLA detection
(sources 1$-$4). The cross marks the average position of all the sources.
{\bf (d)}: Mass segregation ratio parameter $\Lambda_\mathrm{MSR}$ (see text)
as a function of the number of most massive sources.
The black data points and curve show  $\Lambda_\mathrm{MSR}$ calculated 
with sources ordered by their disk masses.
The red data points and curve show $\Lambda_\mathrm{MSR}$ calculated
by setting the four sources with VLA detections (sources 1$-$4) to be the most massive ones 
(ordered by their estimated stellar mass), 
followed by the other sources ordered by their disk masses.
{\color{black} The two outermost sources are not included as the low primary beam response near
the edge of the field of view biases towards relatively bright and therefore massive sources.}
The two sets of data points are offset slightly in x-axis for better presentation.
The dashed line shows $\Lambda_\mathrm{MSR}=1$ indicating no mass segregation.}
\label{fig:dist_func}
\end{center}
\end{figure}

Figure \ref{fig:dist_func}(a) shows disk masses as a function of the projected distance from massive stars.
Here, for any source, we calculate its distances to sources 1 and 2, and use the smaller value as its distance from massive stars.
It appears that the sources within a distance of 3000~au of the massive stars have relatively low disk masses.  
Such a trend can be explained by depletion of disks in the regions close to the massive stars.
It may be due to the tidal/dynamical interactions between the sources in such a crowded region.
In fact, our molecular line observations have revealed complex accretion structures on multiple scales 
that fit such a scenario (see \S\ref{sec:disk}). 
Another possibility of disk depletion close to the massive stars is photoevaporation by high UV radiation. 
However, this is less likely to be the main cause in our case. 
As discussed in \S\ref{sec:spec}, the massive sources in this cluster have not yet developed large {\Hii} regions, 
and are still highly embedded inside dense disk/envelope structures.
However, such a trend may be caused by other effects. 
Many of the sources close to the massive stars are associated with the larger filament and 
the hierarchical accretion structures (see \S\ref{sec:disk}). 
The long-baseline observations detect the small over-densities inside the larger structures associated with
these sources. Whether such over-densities can represent the individual circumstellar disks is uncertain.

Figure \ref{fig:dist_func}(b) shows the distribution function of the projected separations between pairs 
of sources in the cluster. We calculate the distribution function $p(s)$ by counting the number of pairs 
($N_i$) with separation in the interval of $(s,s+\Delta s)$, normalized by the total pair number (e.g., \citealt[]{CW04,Busquet19})
\begin{equation}
p(s)\Delta s= \frac{2N_i}{N_\mathrm{tot}(N_\mathrm{tot}-1)},
\end{equation}
where $N_\mathrm{tot}=22$ is the total source number. 
Here the separation $s$ is normalized to the cluster radius 
{\color{black} $r_\mathrm{max}=18.1\arcsec$  ($4.0\times 10^4$~au),}
which is determined by the maximum separation between any source to the average position of all the sources. 
The distribution function shows a single peak at {\color{black} about 5$\arcsec$ (normalized $s\approx 0.25$).
As demonstrated by \citet[]{CW04}, a single peak in the distribution function is consistent with 
a smooth radial gradient of source density without subclustering.
With $N_\mathrm{tot}=22$ and $r_\mathrm{max}=4.0\times 10^4~\au$, this cluster
has a number density of protostellar sources of $n=7.2\times 10^2~\mathrm{pc}^{-3}$.
There are 20 sources within $2.5\times 10^4~\au$ (primary beam response $>0.5$),
which yields a number density of $n=2.7\times 10^3~\mathrm{pc}^{-3}$.}

We further use the Minimum Spanning Tree (MST) method to characterize the 
clustering of these sources (Figure \ref{fig:dist_func}c). 
It shows the shortest path-length connecting all the sources without including closed loops 
(\citealt[]{Kruskal56}). We construct the MST using the python package NetworkX (\citealt[]{Hagberg08}). 
Following \citet[]{CW04}, from the constructed MST, we derive the $Q$ parameter, which is used to distinguish 
between a smooth overall radial density gradient and multiscale fractal subclustering. 
The $Q$ parameter is defined as $\bar{Q}\equiv \bar{m}/\bar{s}$, 
where $\bar{m}$ is the mean MST path-length normalized by the factor 
$\sqrt{N_\mathrm{tot}A}/(N_\mathrm{tot}-1)$ with $A=\pi r_\mathrm{max}^2$ being the cluster area, 
and $\bar{s}$ is the mean separation of all the sources normalized to the cluster radius. 
{\color{black} From $\bar{m}=0.40$ obtained from MST, and $\bar{s}=0.43$ from the separation distribution
(Figure \ref{fig:dist_func}b), we have $\bar{Q}=\bar{m}/\bar{s}=0.95$.}
$\bar{Q}=0.8$ is considered as a diagnostic boundary; higher values indicate a smooth overall radial 
density gradient and lower values indicate a fractal subclustering (\citealt[]{CW04}). 
{\color{black} The result of $\bar{Q}=\bar{m}/\bar{s}=0.95$ 
is consistent with a cluster without subclustering, 
as the observed single-peaked separation distribution function has suggested.}

Based on the MST method, we further calculate the mass segregation ratio 
$\Lambda_\mathrm{MSR}$ following the definition by \citet[]{Allison09}
\begin{equation}
\Lambda_\mathrm{MSR}=\frac{L_\mathrm{norm}}{L_\mathrm{massive}} \pm 
\frac{\sigma_\mathrm{norm}}{L_\mathrm{massive}},
\end{equation}
where $L_\mathrm{massive}$ is the path-length of the MST of the $N$ most massive sources. 
$L_\mathrm{norm}$ and $\sigma_\mathrm{norm}$ are the average path-length and 
its statistical deviation of the MST of $N$ sources randomly selected from the cluster. 
To estimate $L_\mathrm{norm}$ and $\sigma_\mathrm{norm}$, 
we randomly select 500 sets of sources for each $N$ (\citealt[]{Allison09,Plunkett18}).
{\color{black} Note that we exclude the outermost two sources from this analysis, as
the low primary beam response near
the edge of the field of view strongly biases against dimmer and therefore less massive sources.}

Figure \ref{fig:dist_func}(d) shows $\Lambda_\mathrm{MSR}$ 
as a function of the number of most massive sources in the cluster used in 
constructing the MST analysis ($N_\mathrm{MST}$). 
A value of $\Lambda_\mathrm{MSR}=1$ indicates a random distribution of sources 
without mass segregation, and  $\Lambda_\mathrm{MSR}>1$ suggests that the $N$ most 
massive sources are more concentrated with respect to the random sample, i.e., a mass segregation. 
The black data points show $\Lambda_\mathrm{MSR}$ calculated with sources ordered by their disk masses. 
The mass segregation becomes significant only for the most massive three sources and peaks at 
$N_\mathrm{MST}=3$, which corresponds to sources 1 and 2. 
However, for sources 1$-$4, the level of dust continuum emission is uncertain because 
the estimation of free-free contamination depends on the model (see \S\ref{sec:spec}). 
Meanwhile, these four sources should be the most massive sources as they have already 
formed photoionized regions. The red data points show $\Lambda_\mathrm{MSR}$ 
calculated by setting these four sources to be the most massive ones 
(ordered by their estimated stellar mass), followed by other sources ordered by the estimated disk masses. 
In this case, $\Lambda_\mathrm{MSR}$ clearly has a peak at $N_\mathrm{MST}=4$ indicating that these four 
massive sources show strong segregation from the other members of the cluster. 
{\color{black} No strong mass segregation is seen for $N_\mathrm{MST}\gtrsim 6$, 
which corresponds to $m_\mathrm{disk}<0.16~M_\odot$.}
Considering the short formation time scale of massive stars and the fact that the photoionized regions of 
most massive sources in this cluster are still highly compact, 
the mass segregation is unlikely to be caused by dynamical relaxation 
(as is the case for more evolved clusters). 
However, instead it should reflect the primordial mass segregation during the 
formation of the massive protostellar cluster.

\section{Summary}
\label{sec:summary}

We have presented ALMA 1.3~mm intermediate resolution and long-baseline observations 
towards the massive star-forming region G35.20-0.74 (G35.2).
Using the continuum emission and emission from different molecular lines, we characterized the massive and 
low-mass protostars in the region and their accretion and outflow structures. Our main conclusions are as follows:

(1) The 1.3~mm intermediate-resolution continuum image shows a string of cores along the filamentary structure, 
similar to what has been seen in the previous 0.85~mm observation. 
From the 1.3~mm long-baseline observation, we identified {\color{black} 22} compact continuum sources, 
suggesting a young forming cluster in the region. 
Among them, four sources have corresponding VLA detections from 6 cm to 7 mm.
Three of them (ALMA sources 1, 2 and 3) are consistent with massive young stars which have initiated photoionization,
while the other source (ALMA source 4) can be a low-mass protostar with shock-induced free-free emission. 
{\color{black} Furthermore, we report the detection of a compact source at the location of source 3
in the 0.85~mm continuum, found as a result of reprocessing previously published 0.85~mm continuum data.}
The other {\color{black} 18} sources without corresponding VLA detections are considered to be lower-mass members in the cluster.

(2) Among the massive sources, two (sources 2 and 3) are found to have H30$\alpha$ recombination line emission. 
The H30$\alpha$ line kinematics shows ordered motions of the ionized gas in the innermost region. 
For source 3, the H30$\alpha$ kinematics are consistent with disk rotation, while for source 2 it is more consistent with the outflowing motion. 
We found evidence of potential maser activity in the H30$\alpha$ line emission of source 3,
adding another candidate case to the handful of millimeter HRL maser discovered so far.

(3) We constructed models of photoionized regions to simultaneously fit the multi-wavelength free-free fluxes 
and the H30$\alpha$ total fluxes. The model assumes an isothermal ionized region with a power-law distribution 
in the emission measure (and electron density), so that it can fit the observed SEDs, which show partially optical thick 
free-free emission over wide frequency ranges. The derived properties of the ionized regions are consistent with 
photoionized HC{\Hii} regions.

(4) Molecular line emission shows multi-scale accretion structures around sources 1 and 2 (cores A and B). 
We have inferred that sources 1 and 2 are surrounded by a common rotating structure that connects to the individual disks. 
The source-2 disk is found to have two spiral arms, which may be caused by gravitational instability or perturbation of the binary companion source 4. 
{\color{black} As another possibility, the spiral structures could also be infalling streamers feeding an inner unresolved disk in source 2.}
The spiral arms are best seen in the SO$_2$ emission, which is enhanced by shocks along the spirals. 
SO$_2$ also traces the inner faster-rotating part of the disk while other molecular lines such as CH$_3$OH and H$_2$CO 
only trace the outer slower-rotation part. 
The source 1 disk appears to be more compact than the source 2 disk and not rotationally supported. Its velocity structure
is consistent with infalling-rotation of unsmooth accretion flows.
The existence of such multi-scale accretion structures feeding sources 1 and 2 suggests that 
the accretion is still going on and not stopped by the photoionization feedback.

(5) From the $^{12}$CO emission, we identified in total four outflows likely driven by four different sources 
({\color{black} sources 1, 2, 14, 17}). By comparing with the outflows identified in previous studies with different tracers, 
we speculate that the North-South CO outflow driven by source 2 is consistent with the radio jet and the bright 
outflow cavity seen in NIR and MIR. The wide outflow towards the north-east is actually composed of two separate outflows 
driven by source 1 and {\color{black} source 14}, and it is not formed by the precession of the north-south jet. 
Furthermore, we identify another outflow in the east-west direction, driven by {\color{black} source 17}.

(6) We analyzed the distribution of the identified compact continuum sources in the cluster. 
Assuming the compact dust continuum emission associated with these sources comes from their disks {\color{black} or inner envelopes}, 
we analyzed the relation of their masses with their distances to the massive stars, and 
found potential disk depletion due to dynamical interactions in the inner region. 
The spatial distribution of the sources suggests a smooth overall radial density gradient without subclustering, 
and tentative evidence of primordial mass segregation.


\acknowledgements
This paper makes use of the following ALMA data: 
ADS/JAO.ALMA\#2015.1.01454.S, and
ADS/JAO. ALMA\#2017.1.00181.S.
ALMA is a partnership of ESO (representing its member states), 
NSF (USA) and NINS (Japan), together with NRC (Canada), 
MOST and ASIAA (Taiwan), and KASI (Republic of Korea), 
in cooperation with the Republic of Chile. 
The Joint ALMA Observatory is operated by ESO, AUI/NRAO and NAOJ.
Y.Z. acknowledges support from RIKEN Special Postdoctoral Researcher Program
and JSPS KAKENHI grant JP19K14774.
K.E.I.T. acknowledges support by JSPS KAKENHI Grant Numbers 
JP19K14760, JP19H05080, JP21H00058, JP21H01145.
J.C.T. acknowledges support from ERC project MSTAR, VR grant 2017-04522.
Y.-L. Y. acknowledges the support from the Virginia Initiative of Cosmic Origins (VICO) Postdoctoral Fellowship. 
R.F. acknowledges funding from the European Union’s Horizon 2020 research and innovation 
programme under the Marie Sklodowska-Curie grant agreement No 101032092.
G.G. acknowledges support by the ANID BASAL project FB210003.

\software{CASA (http://casa.nrao.edu; \citealt[]{McMullin07}), 
The IDL Astronomy User's Library (https://idlastro.gsfc.nasa.gov),
NetworkX (https://networkx.org; \citealt[]{Hagberg08})}

\appendix

\section{Model Fitting to the Continuum Spectrum and H30$\alpha$ Emission}
\label{sec:appA}

We construct a simple model to simultaneously fit the continuum flux densities observed in multiple VLA bands and ALMA 1.3~mm, as well as
the total integrated flux density of the H30$\alpha$ line.

The optical depth of the free-free emission (e.g. \citealt[]{Wilson13}) is
\begin{equation}
\tau_{\nu,\mathrm{ff}}  = 8.235\times 10^{-2} \left(\frac{\Te}{\K}\right)^{-1.35}\left(\frac{\nu}{\GHz}\right)^{-2.1}\left(\frac{\elecm}{\pccm}\right). \label{eq:tauff}
\end{equation}
where $\Te$ is the temperature of the ionized gas, and $\elecm$ is the emission measure.
defined as $\elecm\equiv\int n_e^2 dh$.
Assuming that the ionized gas is isothermal, the radiative transfer of the free-free emission gives the brightness temperature as,
\begin{equation}
T_{\nu,\mathrm{ff}} =\Te\left(1-e^{-\tau_{\nu,\mathrm{ff}}}\right).
\end{equation}
The total free-free flux density is then
\begin{equation}
S_{\nu,\mathrm{ff}}=\frac{2k\nu^2}{c^2}\int T_{\nu,\mathrm{ff}}  d\Omega,
\end{equation}
where $\Omega$ is the solid angle.

The optical depth of the HRL line center under local thermal equilibrium (LTE) condition 
{\color{black}
(e.g. \citealt[]{Wilson13}) is}
\begin{equation}
\tau_{\mathrm{HRL},0} = 1.92\times 10^3 \left(\frac{\Te}{\K}\right)^{-5/2}\left(\frac{\elecm}{\pccm}\right)\left(\frac{\dnu}{\kHz}\right)^{-1}.\label{eq:tau_HRL}
\end{equation}
{\color{black}
For H30$\alpha$ line (231.9 GHz), the line center optical depth (Eq. \ref{eq:tau_HRL}) is}
\begin{equation}
\tau_{\mathrm{H}30\alpha,0}  =2.48 \left(\frac{\Te}{\K}\right)^{-5/2}\left(\frac{\elecm}{\pccm}\right)\left(\frac{\dv}{\kms}\right)^{-1},\label{eq:tau_H30alpha}
\end{equation}
where $\dnu$ and $\dv$ are the Full Width at Half Maximum (FWHM) of the line in the frequency and the velocity, respectively.

For a typical electron temperature of $10^4~\K$ and typical HRL line width of $40~\kms$, H30$\alpha$ is optically thin except for extremely high emission measures ($\elecm>1.6\times 10^{11}~\pccm$). Therefore, the brightness temperature of the {\it free-free continuum subtracted} H30$\alpha$ is
\begin{equation}
\Thrl=\Te e^{-\tauff}\left(1-e^{-\tauhrl}\right)=\Te e^{-\tauff} \tauhrl.\label{eq:THRL}
\end{equation}
Note that here the factor of $e^{-\tauff}$ includes the effects of non-optically-thin free-free continuum on the line intensity. 
{\color{black}
From Eqs. \ref{eq:tau_H30alpha} and \ref{eq:THRL}, the integrated H30$\alpha$ intensity is then}
\begin{equation}
\left(\frac{\int T_{\mathrm{H}30\alpha} dv}{\K~\kms} \right) = 2.64 \left(\frac{\Te}{\K}\right)^{-3/2}
\left(\frac{\elecm}{\pccm}\right) e^{-\tauff},
\end{equation}
and the total H30$\alpha$ flux density integrated over velocity and space is
\begin{equation}
S_{\mathrm{H}30\alpha} =  \frac{2k\nu^2}{c^2} \iint T_{\mathrm{H}30\alpha} dv d\Omega.
\end{equation}

Note that if the free-free and HRL emissions are optically thin, we have the line-to-continuum ratio of
\begin{eqnarray}
\left(\frac{\int \Thrl dv}{\Tff} \right)_{\mathrm{thin}} & =  & \left(\frac{\tau_{\mathrm{HRL},0}}{\tauff}\right)
 \dv \left(\frac{1}{2}\sqrt{\frac{\pi}{\ln 2}}\right) \\
 & = & 7.445\times 10^3~\kms \left(\frac{\Te}{\K}\right)^{-1.15}\left(\frac{\nu}{\GHz}\right)^{1.1}.
\end{eqnarray}
For the H30$\alpha$ line, we have
\begin{equation}
\left(\frac{\int T_{\mathrm{H30}\alpha} dv}{\Tff} \right)_{\mathrm{thin}} = 74.7~\kms \left(\frac{\Te}{10^4~\K}\right)^{-1.15}.\label{eq:HRL_thin}
\end{equation}
The true line-to-continuum ratio should be lower than this if the free-free continuum is not optically thin,
or higher than this if non-LTE effects become significant (as would be the case for a HRL maser).

For a uniform distribution of the emission measure,
the free-free spectral index is $\alpha_\mathrm{ff}\approx 2$ (optically thick) in lower $\nu$,
and $\alpha_\mathrm{ff}\approx -0.1$ (optically thin) in higher $\nu$, with a relatively narrow transition
frequency range.
The observed fluxes in VLA bands show consistent spectral indices between 0 and 2 for
a quite wide frequency range, which requires a non-uniform distribution of the emission measure
(and electron density).
For simplicity, we assume the emitting region projected on the plane of sky is 
a circular disk with an outer radius of $r_\mathrm{out}$,
and the emission measure follows a power law dependence on the radius $r$,
\begin{equation}
\elecm(r) =\elecm_{\mathrm{10au}} \left(\frac{r}{10~\au}\right)^{-\rho_\mathrm{EM}}.
\end{equation}
We also assume the disk has an inner radius of $r_\mathrm{in}$ to avoid singularity.
Then we have the free-free and H30$\alpha$ fluxes as
\begin{equation}
\left(\frac{S_{\nu,\mathrm{ff}}}{\Jy}\right) = 4.534 \times 10^{-10} \left(\frac{\nu}{\GHz}\right)^2 \left(\frac{d}{\kpc}\right)^{-2}
\left(\frac{\Te}{\K}\right) \int_{R_\mathrm{in}}^{R_\mathrm{out}}\left(1-e^{-\tauff}\right) R dR, \label{eq:flux_ff}
\end{equation}
and
\begin{equation}
\left(\frac{S_{\mathrm{H}30\alpha}}{\Jy~\kms}\right) = 6.445\times 10^{-5} \left(\frac{d}{\kpc}\right)^{-2}
\left(\frac{\Te}{\K}\right)^{-3/2} \left(\frac{\elecm_{\mathrm{10au}}}{\pccm}\right)
\int_{R_\mathrm{in}}^{R_\mathrm{out}} e^{-\tauff} R^{1-\rho_\mathrm{EM}} dR, \label{eq:flux_HRL}
\end{equation}
where $R=r/10~\au$, $R_\mathrm{in}=r_\mathrm{in}/10~\au$, $R_\mathrm{out}=r_\mathrm{out}/10~\au$,
and $\tauff$ is given by Eq. \ref{eq:tauff}.
These are the observed quantities we fit with the model.

\begin{figure}
\begin{center}
\includegraphics[width=0.8\textwidth]{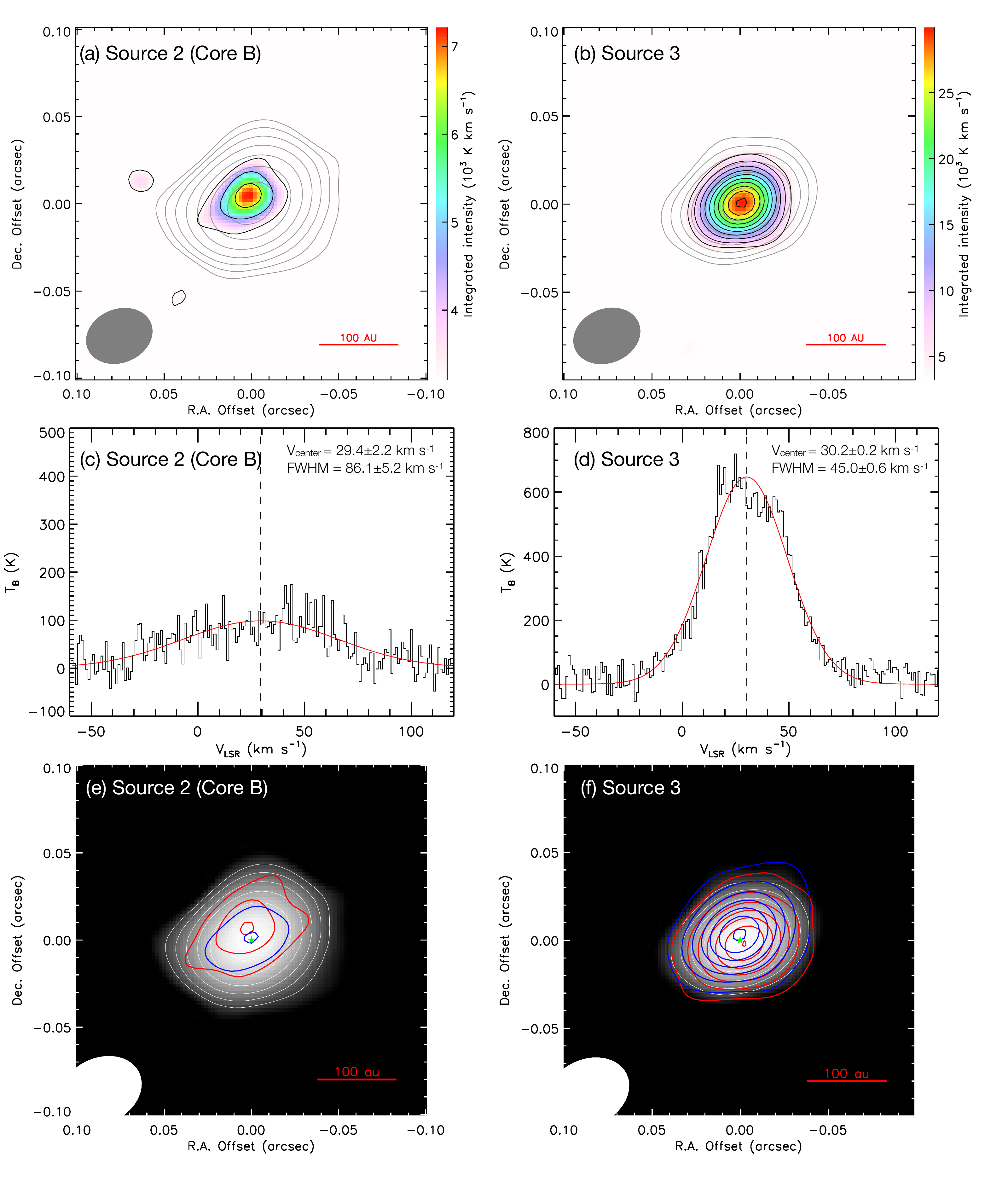}\\
\caption{{\bf (a)}: Integrated emission map of the H30$\alpha$ recombination
line shown in color scale and black contours, overlaid with the continuum emission (grey contours),
for ALMA source 2 (i.e. core B). Both line and continuum images are from the C9 configuration data.
the H30$\alpha$ emission is integrated in the range of $-5~\kms<\vlsr<70~\kms$.
{\color{black} The H30$\alpha$ contours (black) start at 3$\sigma$ 
and have intervals of $1.5\sigma$}, with $1\sigma=0.045~\Jybeam~\kms$ ($5.5\times 10^2~\K$).
The continuum contours (grey) in both panels have levels of $5\sigma\times2^n$ ($n=0,1,...$),
with $1\sigma=0.061~\mJybeam$  ($1.4~\K$).
{\bf (b)}: {\color{black} Same as panel b, but for ALMA source 3, and the H30$\alpha$ contours start at 5$\sigma$ 
and have intervals of $5\sigma$.} 
{\bf (c)}: H30$\alpha$ spectrum at the continuum peak position of ALMA source 2
(black curve), with a Gaussian fit shown by the red curve. The central velocity 
(shown by the vertical dashed line) and FWHM of the
Gaussian fit are labeled.
{\bf (d)}:  Same as panel c, but at the continuum peak position of ALMA source 3.
{\bf (e)}: Integrated blue-shifted and red-shifted H30$\alpha$ emission maps 
(blue and red contours) overlaid on the 1.3~mm continuum emission (grey scale and white contours)
of source 2.
The blue-shifted and red-shifted emission is integrated in the velocity ranges of
$-5~\kms<\vlsr<+30~\kms$ and $+30~\kms<\vlsr<+65~\kms$, respectively.
The blue and red contours have lowest contour 
levels of $5\sigma$ and intervals of $2.5\sigma$,
where $1\sigma=270~\K~\kms$.
The stars mark the continuum peak position.
{\bf (f)}: Same as panel a, but for source 3, and the blue and red contour intervals 
are $5\sigma$.
}
\label{fig:HRL}
\end{center}
\end{figure}

To further reduce the free parameters in the model, we adopt $r_\mathrm{in}=10~R_\odot$ ($0.05~\au$),
and $r_\mathrm{out}=50~\au$.
The inner radius of $10~R_\odot$ is a typical stellar radius for massive protostars reaching main sequence 
(e.g. \citealt[]{ZT18}).
The outer radius of $50~\au$ is roughly consistent with the observation (see Figure \ref{fig:HRL}).
Therefore we have three free parameters in this model,
the electron temperature of the ionized gas ($T_e$), 
the emission measure at the radius of $10~\au$ ($\elecm_{\mathrm{10au}}$),
and the power-law index of the emission measure distribution $\rho_\mathrm{EM}$.
We expect the electron density distribution has a power-law form $n_e \sim r^{-\rho_{n_e}}$ 
with $1.5\leq \rho_{n_e} \leq 2$.
The power-law index of $-1.5$ for the electron density corresponds to the case of
the surface of a photoionized disk around a massive forming star
(\citealt[]{Tanaka13}).
The power-law index of $-2$ of the electron density corresponds to the case 
of an expansion flow of the ionized gas with constant velocity.
The emission measure is then $\elecm\sim n_e^2 (r) l$ 
where $l$ is the line-of-sight length scale of the ionized region.
As a first-order estimate, we assume $l \propto r$, and
$\elecm \sim n_e^2 (r) r \sim r^{1-2\rho_{n_e}}$.
Therefore, we limit the power-law index of the emission measure distribution to the range of 
$2 \leq \rho_\mathrm{EM}\leq 3$ in the fitting.

To find the best-fit model, we compare the model free-free continuum flux densities
(Eq. \ref{eq:flux_ff}) to the continuum fluxes observed in VLA bands,
and compare the model H30$\alpha$ flux integrated over velocity and space to
the observed value from the ALMA observation.
For each model, we calculate the test statistics $\chisq_{\mathrm{ff}}$ and $\chisq_{\mathrm{H}30\alpha}$ as
\begin{equation}
\begin{split}
\chisq_\mathrm{ff}=\sum \frac{(S_{\nu,\mathrm{ff,mod}}-S_{\nu,\mathrm{ff,obs}})^2}{\sigma^2_{\nu,\mathrm{ff,obs}}},\\
\chisq_{\mathrm{H}30\alpha}=\frac{(S_{\mathrm{H}30\alpha,\mathrm{mod}} - 
S_{\mathrm{H}30\alpha,\mathrm{obs}})^2}{\sigma^2_{\mathrm{H}30\alpha,\mathrm{obs}}},
\end{split}
\end{equation}
where the summation of $\chisq_{\mathrm{ff}}$ is for all the VLA bands.
Then, we determine the best-fit model by 
minimizing the value of $\chisq=(\chisq_\mathrm{ff}+\chisq_{\mathrm{H}30\alpha})/N$,
where $N$ is the total number of data points (including continuum and H30$\alpha$ line) used
in fitting.
Note that the ALMA 1.3~mm continuum is not used for fitting the free-free emission, as it usually includes
significant dust emission.

If the H30$\alpha$ line has a non-LTE maser component (such as in source 3), we use the observed
H30$\alpha$ flux as an upper limit.
If the H30$\alpha$ line is not detected (such as in source 1), we use the $3\sigma$ level in
the integrated map as an upper limit.
If the observed H30$\alpha$ and ALMA 1.3~mm continuum fluxes indicate strong dust continuum contributions
even in VLA bands,
we also include the dust component following 
\begin{equation}
S_{\nu,\mathrm{dust}}=S_{\mathrm{234GHz,dust}}\left( \frac{\nu}{234~\GHz} \right)^{\alpha_\mathrm{dust}}, \label{eq:dust}
\end{equation}
and fit all the continuum fluxes (VLA and ALMA) with $S_\nu=S_{\nu,\mathrm{dust}}+S_{\nu,\mathrm{ff}}$.

From the derived properties, we further estimate the ionizing photon rate by balancing the rates
of recombination and photoionization, following the method by \citet[]{Schmiedeke16},
\begin{equation}
\dot{N}_\mathrm{i}=\int n_e^2 (\beta-\beta_1)dV=d^2 \int \elecm (\beta-\beta_1) d\Omega,
\end{equation}
where $\beta$ and $\beta_1$ are the rate coefficients for the recombination to all levels and to the ground state, i.e.,
\begin{equation}
\beta-\beta_1=4.1\times 10^{-10}~\cm^3~\mathrm{s}^{-1} \left(\frac{T_e}{\K}\right)^{-0.8}.
\end{equation}
From our assumptions on the emission measure distribution, we obtain the ionizing photon rate of
\begin{equation}
\dot{N}_\mathrm{i}=1.1 \times 10^{45}~\mathrm{s}^{-1}
\left(\frac{T_e}{10^4~\K}\right)^{-0.8} \left(\frac{\elecm_{\mathrm{10au}}}{10^{10}~\pccm}\right) 
\int_{R_\mathrm{in}}^{R_\mathrm{out}} R^{1-\rho_\mathrm{EM}} dR.\label{eq:ni}
\end{equation}

\clearpage

\begin{table} 
\scriptsize
\begin{center}
\caption{\color{black} Properties of the Presented Continuum Images \label{tab:contimage}}
\begin{tabular}{ccccccc}
\hline
\hline
Telescope & Band &  Wavelength & Configuration & Synthesized Beam & Noise rms & Noise rms \\
& & (mm) &  & & $\mJybeam$ & K \\
\hline
ALMA & Band 7$^{a}$ & 0.85 & extended (cycle 0) & $0\farcs47\times 0\farcs42$ ($\pa=51.4^\circ$) & 0.73 & 0.039 \\
ALMA & Band 6 & 1.3 & C3+C6 & $0\farcs25\times 0\farcs25$ ($\pa=-36.3^\circ$) & 0.48 & 0.17 \\
ALMA & Band 6 & 1.3 & C3+C6+C9 & $0\farcs045\times 0\farcs033$ ($\pa=-64.0^\circ$) &0.085 & 1.3 \\
ALMA & Band 6 & 1.3 & C6+C9 & $0\farcs043\times 0\farcs031$ ($\pa=-63.3^\circ$) & 0.088 & 1.5 \\
ALMA & Band 6 & 1.3 & C9 & $0\farcs037\times 0\farcs027$ ($\pa=-66.0^\circ$) &0.061 & 1.4 \\
VLA & Q band$^{b}$ & 7 & A & $0\farcs046\times 0\farcs036$ ($\pa=-52.0^\circ$) & 0.020 & 7.7 \\
VLA & Q band$^{b}$ & 7 & B & $0\farcs14\times 0\farcs12$ ($\pa=-44.5^\circ$) & 0.030 & 1.1 \\
VLA & Ku band$^{b}$ & 13 & A & $0\farcs079\times 0\farcs076$ ($\pa=-46.3^\circ$) & 0.013& 5.2 \\
VLA & Ku band$^{b}$ & 13 & B & $0\farcs25\times 0\farcs24$ ($\pa=39.5^\circ$) & 0.020 & 0.8 \\
\hline
\end{tabular}
\end{center}
{\bf Notes:} {\bf (a)} {\color{black} re-imaged using the data published by \citet[]{Sanchez13}.} {\bf (b)} \citet[]{Beltran16}. 
\end{table}

\clearpage

\begin{table*} 
\scriptsize
\begin{center}
\caption{Identified compact continuum sources in ALMA 1.3~mm observations \label{tab:cont}}
\begin{tabular}{cccccccccc}
\hline
\hline
\multirow{2}{*}{Source} & VLA & Associated 
& $\alpha$(ICRS) & $\delta$(ICRS) & $I_\mathrm{peak,1.3~mm}$$^{c,d}$ & $S_\mathrm{1.3~mm}$$^{c,e}$ & Mass$^g$ 
& Radius$^h$  & \multirow{2}{*}{Driving Outflow$^i$ }  \\
& Source$^a$ & Core $^b$ & h m s & $^\circ$ $\arcmin$ $\arcsec$ & ($\mJybeam$) & (mJy) & ($M_\odot$) & (au)  & \\
\hline
1 & 3 & A & 18:58:12.953  &  1:40:37.359  & 22.99 $\pm$ 0.06   &   42.19    $\pm$  0.15 (42)$^f$ & 0.70 & 35 & Outflow 2\\
2 & 8a & B & 18:58:13.037  &  1:40:35.931  & 27.80    $\pm$  0.06  &    35.20   $\pm$   0.15 (25)$^f$  &  0.42 & 17 & Outflow 1\\
3 & 1 & & 18:58:12.815  &  1:40:36.604  & 13.03  $\pm$    0.06   &   14.52  $\pm$    0.16 (9.7)$^f$  &  0.16 & $<$12 & \\
4 & 8b & B & 18:58:13.015 &   1:40:36.110  & 1.37   $\pm$   0.06   &    1.49  $\pm$    0.15  (1.3)$^f$  &  0.021 & $<$36 & \\
5 & & &  18:58:12.991  &  1:40:36.200  & 0.74   $\pm$   0.06   &    1.05  $\pm$    0.15 & 0.067  & $<$39 & \\
6 & & &  18:58:12.992  &  1:40:36.338  & 0.75  $\pm$    0.06 &     0.95  $\pm$    0.15 &  0.060 & 17 & \\
7 & & & 18:58:13.070  &  1:40:35.768  &  1.07  $\pm$    0.06   &    1.29  $\pm$    0.15 &  0.083 & 15 & \\
8 & & & 18:58:13.111  &  1:40:35.831  &  0.65   $\pm$   0.06 &    0.65   $\pm$   0.15 &  0.040 & $<$39 & \\
9 & & & 18:58:13.070 &   1:40:37.597  &  1.89  $\pm$    0.06   &   3.65   $\pm$   0.15 &  0.23 & 34 & \\
10 & & & 18:58:12.921  &  1:40:39.988  & 0.86   $\pm$   0.06    &   1.57   $\pm$   0.16 & 0.10& 33 & \\
11 & & D &18:58:12.815  &  1:40:40.000  &  0.89   $\pm$   0.07   &    3.77  $\pm$    0.17 & 0.24 & 257 & \\
12 & & & 18:58:12.836 &   1:40:34.715  & 1.21  $\pm$    0.06  &     2.42  $\pm$    0.16 & 0.15 & 37 & \\
13 & & &18:58:12.725  &  1:40:35.606  &  0.58   $\pm$   0.07  &     0.74  $\pm$    0.17 & 0.047 & 11 & \\
14 & & C &  18:58:13.131 &   1:40:33.233  & 2.98  $\pm$    0.06  &     6.60   $\pm$   0.16 & 0.42 & 44 & Outflow 3 \\
15 & & & 18:58:13.301 &   1:40:34.313  & 0.74   $\pm$   0.07   &    1.88  $\pm$    0.17 & 0.12 & 52 & \\
16 & & & 18:58:12.957  &  1:40:30.656  & 2.12  $\pm$    0.07   &    4.57  $\pm$    0.17 &  0.29  & 38 & \\
17 & & E &  18:58:13.190 &   1:40:30.678  & 0.99   $\pm$   0.07   &    0.99  $\pm$    0.18 &  0.043 & $<$30 & Outflow 4\\
18 & & & 18:58:12.883 &   1:40:31.778  &  1.00    $\pm$  0.07   &    1.48   $\pm$   0.17 &  0.094 & 27 & \\
19 & & &  18:58:12.858  &  1:40:26.403  & 0.77  $\pm$    0.09   &    1.51   $\pm$   0.23 &  0.096 & 34 & \\
20 & & &  18:58:12.290  &  1:40:36.928  &   0.80   $\pm$    0.10   &    1.55   $\pm$    0.25   & 0.099  & 32 & \\
21 & & &  18:58:13.955  &  1:40:27.608  &   1.64   $\pm$   0.20   &    3.14    $\pm$   0.48  & 0.20 & 44 & \\
22 & & &  18:58:12.002  &  1:40:46.487  &   3.05   $\pm$    0.30  &    7.83   $\pm$    0.74   &  0.50 & 57 & \\
\hline
\end{tabular}
\end{center}
{\bf Notes:} {\bf (a)} Compact sources identified from VLA observations by \citet[]{Beltran16}. 
{\bf (b)} Core structures identified from ALMA 0.85~mm observations by \citet[]{Sanchez14}. A source is considered to be
associated with one of the identified cores only if it is close to the continuum peak of that core.
{\bf (c)} Primary beam response has been corrected for the peak intensities, integrated flux densities and their uncertainties.
{\bf (d)} Peak intensity measured from the C9 configuration image.
{\bf (e)} Flux density integrated over a circle with radius of $0\farcs05$. 
The flux density uncertainties are estimated by calculating flux densities with apertures of same size ($0\farcs05$)
over random off-source positions many times on the primary beam uncorrected map, and then scaled
by the primary beam response at that source position.
{\bf (f)} Numbers in the bracket show the dust continuum flux densities, after subtraction of the free-free emission.
The free-free flux densities are estimated from the model fitting to the continuum SED and H30$\alpha$ flux.
{\bf (g)} Gas mass calculated from dust continuum flux densities. A gas-to-dust mass ratio of 100 is assumed.
For sources 1$-$4 (massive sources with photoionized regions), 
a dust temperature of 100 K is assumed, 
and for sources 5$-$22, a dust temperature of 30 K is assumed.
{\bf (h)} Defined as $\sqrt{D_\mathrm{maj}D_\mathrm{min}}/2$, 
where $D_\mathrm{maj}$ and $D_\mathrm{min}$ are the major axis FWHM and minor axis FWHM of the Gaussian
component of the continuum image after deconvolution of the synthesized beam.
The 2D Gaussian fit is performed using the CASA task {\it imfit}.
Upper limits are given if the source is unresolved.
{\bf (i)} Identified from $^{12}$CO maps (see \S\ref{sec:outflow})
\end{table*}

\clearpage

\begin{table*} 
\scriptsize
\begin{center}
\caption{Parameters of the Best-fit Models of Continuum SED and H30$\alpha$ Flux \label{tab:sedmodel}}
\begin{tabular}{cccccccccc}
\hline
\hline
\multirow{2}{*}{Source} & $T_e$ &  $\elecm_\mathrm{10au}$ & \multirow{2}{*}{$\rho_\mathrm{EM}$} 
& $n_{e,\mathrm{10au}}$ & \multirow{2}{*}{$\alpha_\mathrm{dust}$} & 
$\dot{N}_\mathrm{i}$$^a$  & $m_{*,\mathrm{ZAMS}}$ 
& \multirow{2}{*}{spectral type} & $L_{*,\mathrm{ZAMS}}$ \\
& ($10^4$ K) & ($10^9~\pccm$) & & ($10^7~\mathrm{cm}^{-3}$) & & 
($10^{46}~\mathrm{s}^{-1}$) & ($M_\odot$) & & ($10^3~L_\odot$) \\
\hline
source 1 & 1.0 & 0.2 & 2 & 0.2 & 2.3 & $0.009-1$ & $7.3-12.0$ & B2 $-$ B0.5 & $1.8-9.0$ \\
source 2 & 0.9 & 20 & 3 & 2 & & $0.8-48$ & $11.7-18.4$ & B0.5 $-$ O9.5 & $8.4-31$ \\
source 3 & 0.6 & 6.3 & 3 & 1.1 & & $0.3-30$ & $10.7-17.3$ & B1 $-$ B0 & $6.2-26$ \\
source 4 & 1.2 & 0.2 & 2 & 0.2 & & 0.013 & 7.6 & B2 & 2.1 \\
\hline
\end{tabular}
\end{center}
{\bf (a)} As $\dot{N}_\mathrm{i}$ is sensitive to $\rho_\mathrm{EM}$, we show
a range of $\dot{N}_\mathrm{i}$ if  $\rho_\mathrm{EM}$ is
not well constrained by the fitting (see text for details).
\end{table*}

\clearpage

\begin{table} 
\scriptsize
\begin{center}
\caption{Parameters of the Observed Lines\footnote{Molecular line information taken from the 
CDMS database (\citealt[]{Muller05})} \label{tab:lines}}
\begin{tabular}{lccccccc}
\hline
\hline
Species & Transition & Frequency & $E_u/k$ & $S\mu^2$ & Velocity Resolution & Synthesized Beam & Channel rms\\
 & & (GHz) & (K) & (D$^2$) & ($\kms$) & & ($\mJybeam$)\\
\hline
H & 30$\alpha$ & 231.90090 &  & &  0.63 & $0.039\arcsec\times 0.031\arcsec$ ($\pa=-68.0^\circ$) & 1.7 \\
\multirow{2}{*}{SO$_{2}$} &  \multirow{2}{*}{$22_{2,20}-22_{1,21}$} & \multirow{2}{*}{216.6433035} & \multirow{2}{*}{248} & \multirow{2}{*}{35.3} & \multirow{2}{*}{0.68} & $0.30\arcsec\times 0.29\arcsec$  ($\pa=-3.5^\circ$) & 3.3 \\
& &  & & & & $0.052\arcsec\times 0.040\arcsec$  ($\pa=-63.9^\circ$) & 1.9 \\
\multirow{2}{*}{CH$_{3}$OH} & \multirow{2}{*}{$4_{2,2}-3_{1,2}$; E} & \multirow{2}{*}{218.4400630} & \multirow{2}{*}{45.5} & \multirow{2}{*}{13.9} & \multirow{2}{*}{0.68} & $0.29\arcsec\times 0.28\arcsec$  ($\pa=-5.9^\circ$) & 5.6 \\
& & & & & & $0.052\arcsec\times 0.040\arcsec$  ($\pa=-64.3^\circ$) & 2.0 \\
H$_{2}$CO & $3_{2,1} - 2_{2,0}$ & 218.7600660 & 68.1 & 9.06 & 0.67 & $0.29\arcsec\times 0.28\arcsec$  ($\pa=-3.7^\circ$) & 5.4 \\
H$_{2}$S & $2_{2,0} - 2_{1,1}$ & 216.7104365 & 84.0 & 2.06 & 0.68 & $0.30\arcsec\times 0.29\arcsec$  ($\pa=-4.9^\circ$) & 3.4 \\
SiO & $5-4$ & 217.1049190 & 31.3 & 48.0 & 0.67 & $0.30\arcsec\times 0.29\arcsec$  ($\pa=-8.6^\circ$) & 2.9 \\
\multirow{4}{*}{CH$_{3}$OCH$_3$} & $13_{0,13}-12_{1,12}$; AA & 231.9877830 &  69.8  & 170 & 
\multirow{4}{*}{0.63} & \multirow{4}{*}{$0.28\arcsec\times 0.27\arcsec$  ($\pa=-36.5^\circ$)} & \multirow{4}{*}{2.9} \\
 & $13_{0,13}-12_{1,12}$; EE & 231.9878580 &  69.8 & 272 & & & \\
 & $13_{0,13}-12_{1,12}$; AE & 231.9879320 &  69.8  & 102 & & & \\
 & $13_{0,13}-12_{1,12}$; EA & 231.9879330 &  69.8  & 68.0 & & & \\
$^{12}$CO & $2-1$ & 230.5380000 & 16.6 & 0.0242 & 0.63 & $0.28\arcsec\times 0.27\arcsec$  ($\pa=-14.1^\circ$) & 4.0 \\
\hline
\end{tabular}
\end{center}
\end{table}

\clearpage

\begin{table} 
\scriptsize
\begin{center}
\caption{Parameters of the Best-fit Spirals \label{tab:spiral}}
\begin{tabular}{l|cccc|cccc}
\hline
\hline
\multirow{3}{*}{Spiral Arm} & \multicolumn{4}{c|}{Logarithmic ($R=R_0 e^{b\theta}$)} &  \multicolumn{4}{c}{Archimedean ($R=R_0+b\theta$) } \\
& $R_0$ & $b$ & Pitch Angle & $\chi^2$ & $R_0$ & $b$ & Pitch Angle\footnote{Pitch angle from $0.2\arcsec$ to $0.8\arcsec$.} & $\chi^2$ \\
& (arcsec) & (deg$^{-1}$)  & (deg) & & (arcsec) & (arcsec deg$^{-1}$) & (deg) & \\
\hline
NW & 0.042 & 0.044 & 69$^\circ$ & 0.77 & -0.69 & 0.022 & $81^\circ-57^\circ$ & 57 \\
SE & 0.00032 & 0.032 &  62$^\circ$ & 3.9 & -2.96 & 0.015 & $78^\circ -48^\circ$ & 48 \\
\hline
\end{tabular}
\end{center}
\end{table}

\clearpage

\clearpage

\clearpage

\clearpage

\clearpage

\clearpage

\clearpage


\clearpage

\clearpage

\clearpage

\clearpage

\clearpage

\clearpage

\clearpage

\clearpage

\clearpage

%
%

\clearpage

%
%

\clearpage

\clearpage

\clearpage

%

\clearpage

%

\end{document}